\documentclass[11pt,a4paper]{article}
\usepackage{epsfig}
\usepackage{diagbox}
\usepackage{graphicx}
\usepackage{amsmath}
\usepackage{amssymb}
\usepackage{verbatim}
\usepackage{bm}
\usepackage{dsfont}
\usepackage[footnotesize]{caption}
\usepackage{subfigure}
\usepackage{cite}
\usepackage{textcomp}
\usepackage{calc}
\usepackage{geometry}
\usepackage{bbm}
\usepackage{xcolor}
\usepackage[utf8]{inputenc}
\def\ba{\begin{eqnarray}}
\def\ea{\end{eqnarray}}
\usepackage{ulem}
\usepackage{feynmp}
\usepackage{tikz-feynman}
\tikzfeynmanset{compat=1.1.0}

\usepackage{float}
\usepackage{soul}

\usepackage{hyperref}
\hypersetup{colorlinks=true,urlcolor=blue,linkcolor=red,citecolor=green!60!black}

\renewcommand{\baselinestretch}{1.3}

\newcommand{\be}{\begin{equation}}
\newcommand{\ee}{\end{equation}}
\newcommand{\beq}{\begin{equation}}
\newcommand{\eeq}{\end{equation}}
\newcommand{\bea}{\begin{eqnarray}}
\newcommand{\eea}{\end{eqnarray}}

\newcommand{\mgamma}{m_{\gamma'}}
\newcommand{\mphi}{m_\phi}
\newcommand{\opl}{\omega_{pl}}
\newcommand{\og}{\omega_{\gamma'}}

\newcommand{\Gg}{g_{\phi\gamma\gamma'}}
\newcommand{\epdm}{{\bf E}'_{dm}}

\def\d{\delta}
\def\e{\epsilon}           % Also, \varepsilon
\def\f{\phi}               %      \varphi

\def\g{\gamma}

\def\k{\kappa}

\def\o{\omega}

\def\6{\partial}

\def\OO{\Omega_{\bf k}}
\def\sk{s_{\bf k}}

\begin{document}

%opening
\begin{center}

%\bigskip
\vspace*{2.5cm}
{\Large{\textbf{
Hidden Photon Dark Matter Interacting via Axion-like Particles
}}}

\vspace*{1.2cm}

{\large
Paola Arias$^{a,\, b}$,\,%
Ariel Arza$^{c}$,\,%
Joerg Jaeckel$^{d}$\,%
and Diego Vargas-Arancibia$^{a}$%%
}\\[3mm]
{\it{
$^{a}$ Departamento de Fisica, Universidad de Santiago de Chile, Casilla 307, Santiago, Chile
\\
$^{b}$ 
AstroCeNT, Nicolaus Copernicus Astronomical Center Polish Academy of Sciences,\\
ul. Rektorska 4, 00-614 Warsaw, Poland\\
$^{c}$ Institute for Theoretical and Mathematical Physics, Lomonosov Moscow State University (ITMP), 119991 Moscow, Russia\\
$^{d}$ Institut f\"ur theoretische Physik, Universit\"at Heidelberg, \\Philosophenweg 16, 69120 Heidelberg, Germany
}}

\end{center}
\vspace*{1cm}

\begin{abstract}
We investigate a scenario where the dark matter of the Universe is made from very light hidden photons transforming under a $Z_{2}$-symmetry. In contrast to the usual situation, kinetic mixing is forbidden by the symmetry and the dark photon interacts with the Standard Model photon only via an axion-like particle acting as a ``messenger''. Focusing on signatures involving the ordinary photon, our survey of the phenomenology includes limits from cosmological stability, CMB distortions, astrophysical energy loss, light-shining-through-walls experiments, helioscopes and solar X-ray observations.
\end{abstract}

%%%%%%%%%%%%%%%

\newpage

{\hypersetup{linkcolor=black}\renewcommand{\baselinestretch}{1}\tableofcontents}

%%%%%%%%%%%%%%%%%%%

\section{Introduction}
Very light bosons such as axions, axion-like particles (ALPs) and hidden photons (HPs) are amongst the most minimal extensions of the Standard Model allowing for an explanation of the observed dark matter (DM). Their sufficient production can proceed, for example, via the misalignment mechanism~\cite{Abbott:1982af,Preskill:1982cy,Dine:1982ah,Nelson:2011sf,Arias:2012az}\footnote{Other interesting production mechanisms include, the generation from inflationary perturbations (e.g.~\cite{Graham:2015rva,Cosme:2018nly,Alonso-Alvarez:2018tus,Tenkanen:2019aij,AlonsoAlvarez:2019cgw,Ema:2019yrd,Ahmed:2020fhc}), but also from couplings to other (often misaligned) fields that can produce the desired particles in connection with a parametric resonance (see~\cite{Agrawal:2018vin,Co:2018lka,Dror:2018pdh,Bastero-Gil:2018uel} for some examples). Topological defects forming during a phase transition, e.g. strings and domain walls, are also a potentially abundant source of light bosons (cf., e.g.,~\cite{Sikivie:1982qv,Davis:1986xc,Harari:1987us,Davis:1989nj,Dabholkar:1989ju,Hagmann:1990mj,Battye:1993jv,Battye:1994au,Ringwald:2015dsf,Long:2019lwl}).}  and does not require any messenger particles. Similarly, no symmetry\footnote{For a situation where there is an approximately conserved charge and even a charge asymmetry for a very light boson, see, e.g.~\cite{Alonso-Alvarez:2019pfe}.} is required for them to achieve cosmological stability as the decay rate is very suppressed by a combination of very small mass and very weak coupling.

Another reason for their popularity is certainly that these particles lead to new possibilities for their detection in experiments (see, e.g.~\cite{Jaeckel:2010ni,Graham:2015ouw,Irastorza:2018dyq,Sikivie:2020zpn} for some reviews).
In most simple models of such very light dark matter, the dark matter particle itself couples directly -- albeit very very weakly -- with the Standard Model particles, without involving any additional messengers.
For example, in the case of hidden photons~\cite{Okun:1982xi,Holdom:1985ag,Foot:1991kb} (for a review and more literature see, e.g.~\cite{Jaeckel:2013ija}), interactions to the Standard Model (SM) proceed through a kinetic mixing term~\cite{Holdom:1985ag,Foot:1991kb} $\sim\chi F_{\mu\nu}F'^{\mu\nu}$, where $F$ is the electromagnetic field strength of the standard model and $F'$ the corresponding one in the hidden sector. Similarly, for axions and axion-like particles there is a direct interaction with photons (the most exploited coupling) through $\sim \phi F_{\mu\nu}\tilde F^{\mu\nu}$, where $\phi$ is the ALP field.

In contrast many models of WIMP dark matter feature interactions that proceed through additional ``messenger'' particles with masses smaller or comparable to those of the dark matter particles. It is therefore a viable question, what if very light particles interact with the Standard Model only via interactions involving additional particles?

In this work we want to study the phenomenology of a such a model.
Concretely we investigate a situation where the dark matter is made from hidden photons.
However, instead of directly interacting via kinetic mixing the hidden photons couple to the ordinary photon only with the involvement of an axion-like particle via an interaction
$\sim \phi F^{'}_{\mu\nu}\tilde F^{\mu\nu}$.

While other papers have investigated similar models involving axions and hidden photons, they mostly consider different perspectives and situations.
For example in~\cite{Jaeckel:2014qea} such a model was used to explain the 3.5~keV line~\cite{Bulbul:2014sua,Boyarsky:2014jta}, whereas~\cite{Gao:2020wer} discusses such a situation in the context of the Xenon1T result~\cite{Aprile:2020tmw}.
In~\cite{Kalashev:2018bra} this setup was used to address an excess in the cosmic infrared background~\cite{Kohri:2017oqn,Matsuura:2017lub} but they also investigated astrophysical constraints that we will return to later in Sect.~\ref{sec:solar}. In addition, such a coupling can also be useful to generate magnetic fields in the Universe~\cite{Choi:2018dqr}. Refs.~\cite{Co:2018lka, Agrawal:2018vin,Bastero-Gil:2018uel} find that the coupling $g_{\phi\gamma'\gamma'}\phi F_{\mu\nu}'\tilde F'^{\mu\nu}$ could cause axion dark matter (produced through the misalignment mechanism) to covert into hidden photon dark matter for a suitable range in parameter space. Moreover,~\cite{Agrawal:2017eqm,Kitajima:2017peg,Hook:2019hdk} investigated a coupling of the QCD axion to a massless hidden photon through an aligned mechanism, 
finding that the energy transfer between them can open up the parameter space for axions to larger values of $f_{a}$.

Phenomenologically the interaction of axion-like particles with a hidden photon background allows for interesting new possibilities.
In particular, a hidden photon DM condensate
introduces a time-dependent background, where novel effects can take place~\cite{Espriu:2011vj,Espriu:2014lma,Arias:2016zqu,Arza:2017phd,Arza:2017uqa}. Moreover, it provides an environment where axions and photons can couple in ``vacuum'', which can lead to new effects in laboratory experiments.

Our approach will be to first  look into the viability of the model, addressing for instance the stability of the DM, and whether it survives current stringent observations such as constraints on CMB distortions or the solar luminosity.
On the other hand, we would like to test the model with laboratory experiments, and search for distinctive features that could differentiate it from the already known one-particle models, such as axion-like particles or hidden photons.

Let us now briefly outline the structure for the rest of the paper. In section \ref{sec:model} we provide details of the model and its interactions. We also discuss the equations of motion and comment on some of the approximations we employ. In section \ref{sec: stability} we investigate the stability of the dark matter, taking into account the main decay channels, both spontaneous and stimulated, and estimate a bound on the coupling constant based on parametric decay.
In section \ref{sec:CMB} we use the CMB to constrain our model, we consider that the photons produced from the decay $\gamma'\rightarrow \phi+\gamma$ or those evaporated by the annihilation $\gamma'+\gamma\rightarrow\phi$ could distort the blackbody spectrum. In section \ref{sec:solar} we use constraints on an excess energy loss in the sun and in horizontal branch stars to obtain limits on the couplings.

\begin{figure}[t]
    \centering
    \includegraphics[width=0.65\textwidth]{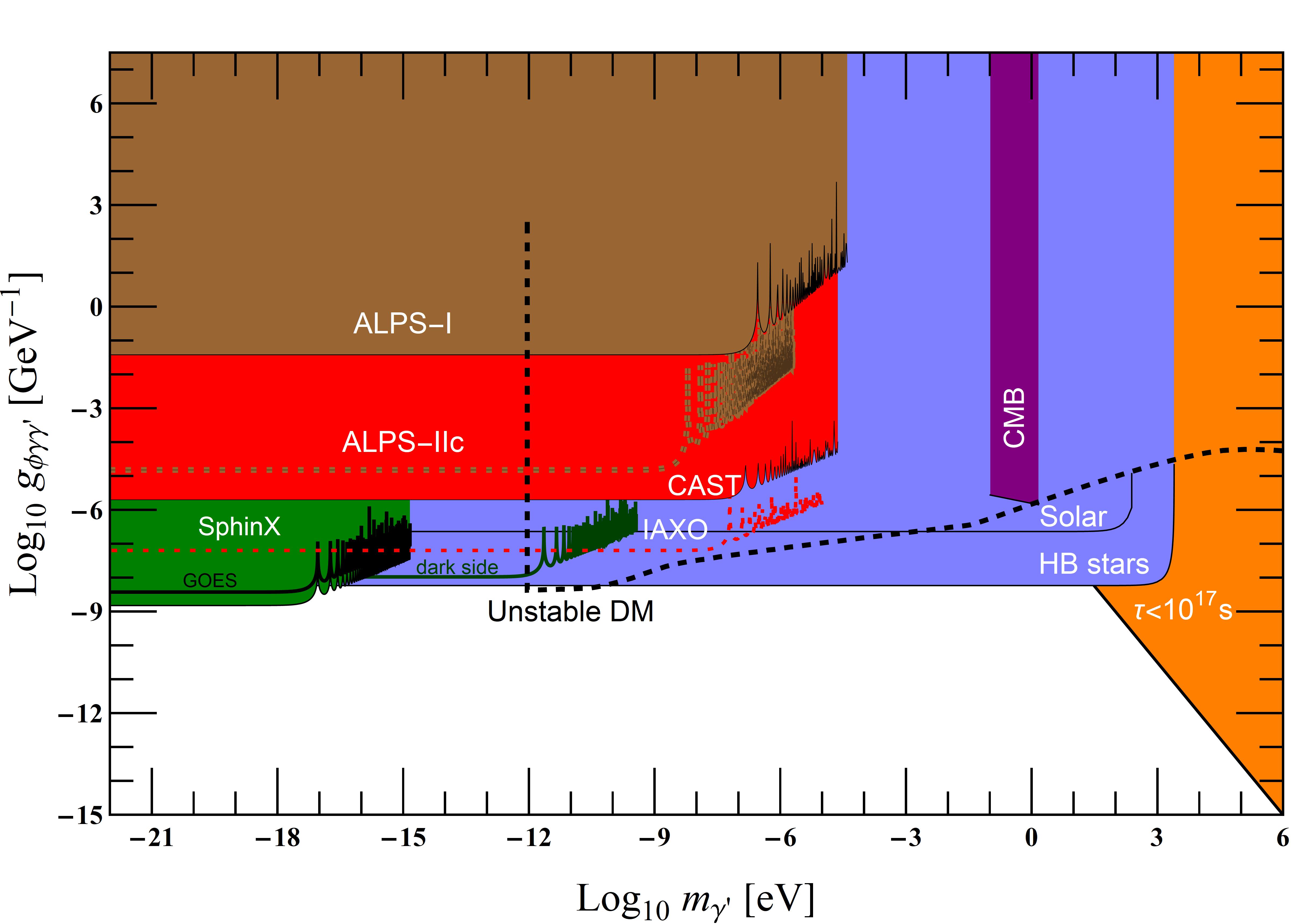}
   \caption{ Summary of the parameter space for the axion-like particle--hidden photon (ALP--HP) coupling as a function of the HP mass as analysed in this work for massless ALPs. Above the black dashed line the dark matter decays via a parametric resonance into a photon and an ALP (cf. Sect.~\ref{sec: stability}). The orange region labeled $\tau<10^{17}$~s is ruled out from the spontaneous decay $\gamma'\rightarrow \gamma+\phi$, discussed in Sect.~\ref{sec: stability}. The purple region labeled ''CMB" is excluded by limits on the CMB distortions that would be induced by decay/evaporation of the DM caused by CMB photons, as discussed in Sect.~\ref{sec:CMB}. The blue regions labeled ''Solar" and ''HB stars'' are derived from the extra energy loss in these objects, see Sect.~\ref{sec:solar}. The brown regions labeled ''ALPS-I"~\cite{ehret2010new} and ``ALPS-II''~\cite{Bahre:2013ywa,Graham:2015ouw} are from the light-shining-through-walls experiments discussed in Sect.~\ref{sec:laboratory}. The red region corresponds to the limits from CAST helioscope~\cite{Anastassopoulos:2017ftl} and the projection for IAXO~\cite{Armengaud:2014gea} (Sect.~\ref{sec:laboratory}). Finally, the green region and lines depict the results from X-ray observations of the sun, discussed at the end of Sect.~\ref{sec:laboratory}. We note that all limits except for the Solar and HB stars constraint are based on the assumption that HPs are the DM. We have chosen to show the case where the DM polarisation is randomly oriented in space and we only show regions where the coherence is not lost due to the effects of structure formation (cf.~Appendix~\ref{sec:coherence}).}
    \label{fig:overview_massless}
\end{figure}

\begin{figure}[t]
    \centering
    \includegraphics[width=0.65\textwidth]{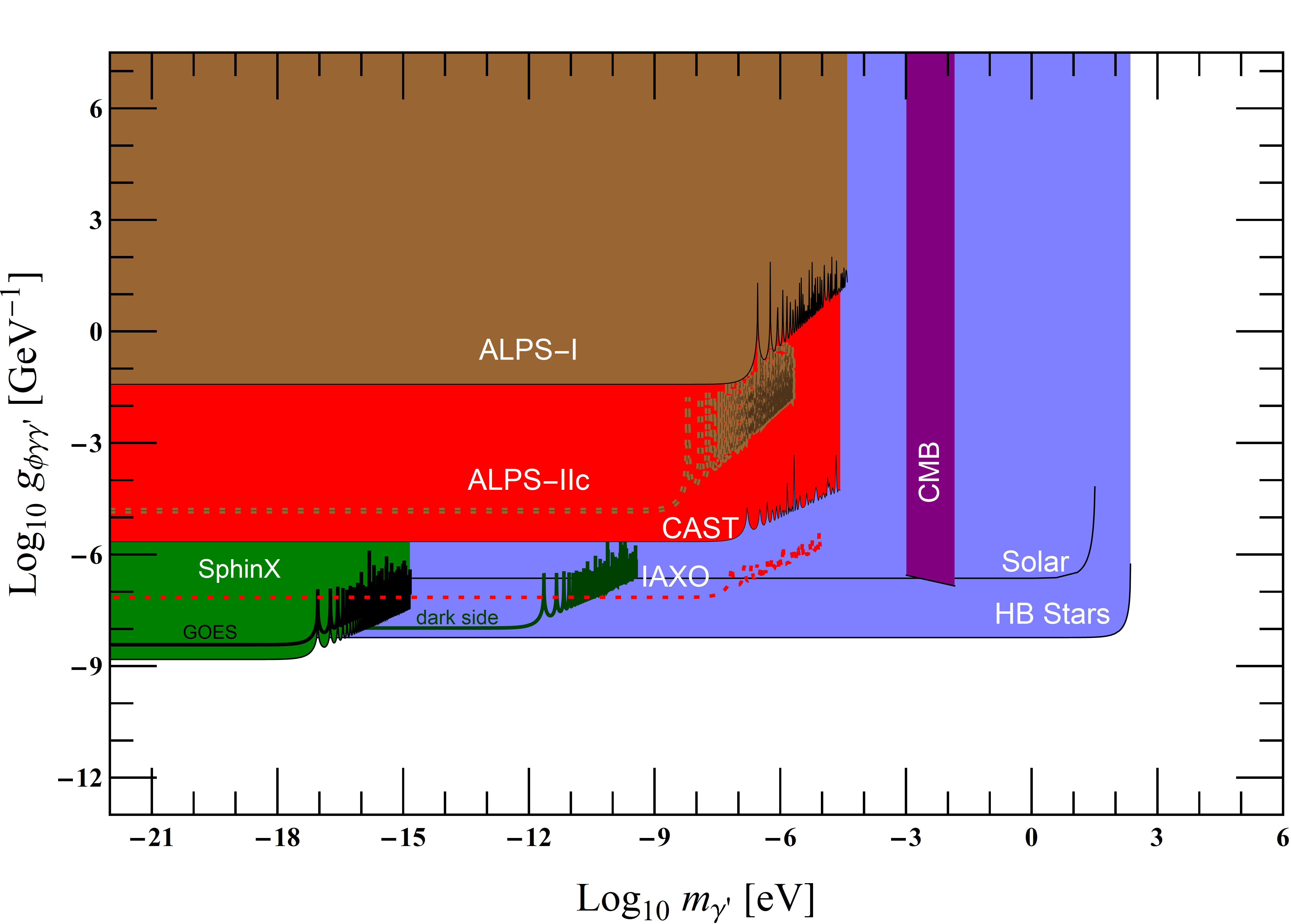}
   \caption{Summary of the parameter space for the axion-like particle--hidden photon (ALP--HP) coupling as a function of the HP mass as analysed in this work  for a mass of the ALP of $\mphi=10\,\mgamma$, so  the DM is stable. All other choices about the considered scenario as well as the colours for the different bounds are the same as in Fig.~\ref{fig:overview_massless}. }
   \label{fig:overview_10mhp}
   \end{figure}

In section \ref{sec:laboratory} we look into laboratory bounds for our model, analyzing  the conversion probability of photons into axions, by interacting with a hidden-photon DM background. We set constraints using light-shining-through-walls (LSW) results. We also very briefly discuss optical effects, such as birefringence and dichroism. At the end of section~\ref{sec:laboratory} we then also combine the flux from the sun with the conversion probability of LSW setups to obtain limits via helioscopes and space X-ray observations\footnote{We thank Gonzalo Alonso-~\'Alvarez for pointing out this process, see also footnote~\ref{foot:thanks}.}. Fig.~\ref{fig:overview_massless} and \ref{fig:overview_10mhp} summarize all these results.
Finally, section~\ref{sec:conclusions} contains a brief summary and conclusions.

%%%%%%%%%%%%%%%%%%%%%%%%%%%%%

\section{The model}{\label{sec:model}}

As discussed in the introduction, we would like to study a model of axion-like particles coupled to hidden photons. Such a model combines two of the most popular models for very light particles. 
As already mentioned aspects of the phenomenology of such a system have already been studied in~\cite{Jaeckel:2014qea,Alvarez:2017eoe,Co:2018lka,Agrawal:2018vin,Agrawal:2017eqm,Gao:2020wer, Nakayama:2020rka}.
Interesting theoretical motivation can be found, for example, in~\cite{Kaneta:2016wvf} where they show it is possible to account for such a coupling to the axion without spoiling the solution of the strong CP problem. 

Following the motivation laid out in the introduction we are specifically interested in a situation where the dark matter interacts with the Standard Model particles only under involvement of an additional particle. Therefore we 1) Need either the axion-like particle or the hidden photon to be the dark matter 2) We need to restrict the interactions such that any interaction with Standard Model particles requires the involvement of more than one dark sector particle.

For 1) We choose the dark matter to dominantly consist of hidden photons. This choice can be motivated from a possible production via the misalignment mechanism~\cite{Nelson:2011sf,Arias:2012az,AlonsoAlvarez:2019cgw}, (resonant) decays of a precursor field~\cite{Agrawal:2018vin,Co:2018lka,Dror:2018pdh,Bastero-Gil:2018uel}, quantum fluctuations grown during inflation~\cite{Graham:2015rva,AlonsoAlvarez:2019cgw,Ema:2019yrd,Ahmed:2020fhc}, energy transfer from the inflaton itself through a kinetic coupling \cite{Nakayama:2019rhg, Nakai:2020cfw}, or the decay of topological defects~\cite{Long:2019lwl}. Ultimately this is, however, a choice we make. The alternative case of the ALPs being the dominant form of dark matter is left for future work.

Requirement 2), needs us to restrict the possible {interaction terms.} In particular the most general system of interactions of HPs includes a kinetic mixing term with the ordinary photon,
\begin{equation}
    {\mathcal{L}}\supset -\frac{1}{2}\chi F'_{\mu\nu}F^{\mu\nu}.
\end{equation}
This term does not involve the ALP field and we therefore want to prevent it.
To remedy this we impose an unbroken $Z_{2}$ symmetry,
\begin{eqnarray}
A^{'}_{\mu}&\rightarrow& -A^{'}_{\mu}\\\nonumber
\phi&\rightarrow& -\phi\\\nonumber
{\rm SM}&\rightarrow&{\rm SM}. 
\end{eqnarray}
Where $A^{\mu}$ and $\phi$ are the HP and ALP field, respectively, and SM indicates
any Standard Model particle.

The Lagrangian  invariant  under the $Z_{2}$ symmetry and taking into account operators of at most mass dimension 5, then reads,
\be \label{lagrangian}
\mathcal L=-\frac{1}4 F_{\mu\nu}F^{\mu\nu}-\frac{1}4 F'_{\mu\nu}F'^{\mu\nu}+\frac{\mgamma^2}2A'_\mu A'^\mu+\frac{1}2\partial_\mu\phi\partial^\mu\phi-\frac{\mphi^2}2\phi^2+\frac{g_{\phi\gamma\gamma'}}{2}\phi F_{\mu\nu}\tilde F'^{\mu\nu}.
\ee
As we can read off, the only interaction between the new fields and the SM photon involves both the ALP and the HP.

In principle we could also include a Higgs-portal term for $\phi$. However, if we take $\phi$ to be a pseudo-Goldstone boson this is likely to be suppressed by the corresponding shift symmetry. Furthermore, it would mostly modify the interactions of $\phi$ and not that of our dark matter particle $A'$. Finally, we also note that such symmetric Higgs portal couplings are generally not very well constrained (cf., e.g.,~\cite{Arcadi:2019lka} for the standard constraint due to invisible Higgs decays valid at low masses and~\cite{Bauer:2020nld} for a somewhat better constraint from supernova cooling\footnote{{See~\cite{Noble:2007kk,Curtin:2014jma,Craig:2014lda,Englert:2020gcp} for a discussion of even weaker constraints at masses above $m_{h}/2$.}}) . Much stronger constraints arise if the ALPs were the DM~\cite{Hees:2018fpg,Bauer:2020nld}. In any case, in the following we will not consider a Higgs portal coupling.

\bigskip

For convenience let us provide the relevant equations of motion for the system in the hidden photon background. In the following sections and in the appendices (where more details can be found) we then solve them for various situations and boundary conditions.
To simplify the equations of motion, we use that the Lorenz condition on the hidden photon background is a consequence of the equations of motion. We therefore have, $\partial_t A'_0=-\nabla\cdot \bf A'$.  
Consequently, the $A'_0$ part is suppressed by the hidden photon velocity and we will neglect it in the following. For the ordinary photon field we use\footnote{Here it is really a choice.} Lorenz gauge $\partial_\mu A^\mu=0$ {as well as} $A_0=0$.  

In the situations we are studying we will use the dark matter hidden photon field as a background and do not explicitly treat backreactions on it.

Allowing for a hidden photon dark matter background and linearising the equations by assuming terms such as $\propto \phi {\bf A}', \phi {\bf A}$ are small we then obtain relatively simple equations for the photon and the ALP field,
\bea
\label{eq:eom}
\left(\partial_t^2-\nabla^2\right){\bf A}&=&-\Gg \nabla\phi \times {\bf E}'_{dm} \label{eom_1}\\
\left(\partial_t^2-\nabla^2+\mphi^2\right)\phi&=&-\Gg \, {\bf E}'_{dm}\cdot {\bf B} , \label{eom_2}
\eea
where $\epdm$ denotes the hidden electric field of the hidden photon dark matter.
At our level of approximation the dark matter hidden photon field obeys 
$\left(\square+\mgamma^2 \right){\bf A}'_{dm}=0$ and it is related to the local dark matter density via~\cite{Arias:2012az,Horns:2012jf},
\begin{equation}
    |\epdm|=|\partial_{t}{\bf A}'_{dm}|=m_{\gamma'}|{\bf A}'_{dm}|=\sqrt{2\rho_{\rm{CDM}}}=3\times 10^3\frac{\rm V}{\rm m}\left(\frac{\rho_{\rm CDM}}{300\,{\rm MeV}/{\rm cm}^3}\right)^{1/2}.
\end{equation}

As we can see from the equations of motion the background of HP-DM leads to an interaction between axions and photons. For a fixed direction of the HP field in space, it is the component of the photon perpendicular to the DM-electric field that mixes with the axion, with an effective strength given by $\Gg \sin\theta$, where $\theta$ is the angle between the dark matter polarisation and the direction of propagation of the incoming photon (see Appendix~\ref{app:eoms} for details). At this point, it is therefore important to briefly comment on the HP polarisation. At present it is not known whether the dark matter condensate has a specific polarisation in space or is randomly oriented (see also~\cite{Arias:2012az}). Therefore, for now we will assume the hidden dark matter electric field can be written as ${\bf E}'_{dm}=E_0'\cos(\mgamma t) \hat \varepsilon_{dm}$, where the energy of the DM particles is $\mgamma+\mathcal O(\mgamma v^2)$. When needed for the bounds from observations and experiments, we will comment on how to treat the DM polarisation.  

Another, perhaps even more important aspect is the question of coherence.
Above we have approximated the HP DM field as constant in space. This is equivalent to assuming that the dark matter is at rest (in the chosen frame). While DM particles are generally slow, over longer length scales their velocity (spread)
cannot be neglected as it leads to a loss of coherence. At this point we expect that our constant field approximation breaks down. The size of density and velocity fluctuations changes during the evolution of the Universe and therefore so will the coherence length. We discuss this in more detail in Appendix~\ref{sec:coherence}. 
Moreover, the DM velocity also depends on the production mechanism. In the case of pure misalignment production~\cite{Nelson:2011sf,Arias:2012az,AlonsoAlvarez:2019cgw} with a field homogenized by inflation the initial velocities are negligible. However, if dark matter is produced from fluctuations or decays~\cite{Agrawal:2018vin,Co:2018lka,Dror:2018pdh,Bastero-Gil:2018uel,Graham:2015rva,AlonsoAlvarez:2019cgw,Ema:2019yrd,Ahmed:2020fhc,Long:2019lwl}, the velocity spread in the early Universe may be significantly larger than that imposed from structure formation. This weakens the cosmological limits discussed in Sect.~\ref{sec: stability} and \ref{sec:CMB}. Those should therefore be considered to be strictly valid only in the case of a sufficiently homogeneous/cold production such as from the misalignment mechanism. For other production mechanisms they have to be studied on a case by case basis. Appendix~\ref{app:productioncoherence} gives some estimates relevant to these situations.
In the following, unless stated otherwise, we assume that the velocities imprinted during production are negligible.
More details on the requirements on the coherence in a given experimental or observational setting are discussed in context in the respective sections.

%%%%%%%%%%%%%%%%%%%%%%%%

\section{Stability of the {Dark Matter}}\label{sec: stability}

One of the primary tests that any dark matter candidate must pass is that it sufficiently long-lived.
In our model we have two distinct regimes,
\begin{equation}
\label{eq:stability}
\begin{array}{ccl}
1)\,\,m_{\gamma'}\leq m_{\phi} &\Rightarrow & \gamma'\,\,\,{\rm is}\,\,\,{\rm stable}
\\
2)\,\,m_{\gamma'}>m_{\phi} &\Rightarrow &\gamma'\,\,\,{\rm can}\,\,\, {\rm decay}\,\,\,{\rm but}\,\,\,{\rm may}\,\,\,{\rm be}\,\,\,{\rm long-lived}.
\end{array}
\end{equation}
The first case is simple, as the stability is ensured by the unbroken $Z_{2}$ symmetry of our model under which the hidden photon is the lightest charged particle\footnote{In principle one could imagine that due to the relatively high occupation numbers of our light dark matter bosons processes with more than one $\gamma'$ may be possible. However, this is a relatively small effect as it involves higher powers of the small coupling.}.

Let us therefore turn to the second possibility of Eq.~\eqref{eq:stability}.
To study the stability of the homogeneous, oscillating field we start by solving the equations of motion~\eqref{eq:eom} from the previous section in the rotating wave approximation\footnote{For our purposes this is essentially equivalent to using a comparison with the Mathieu equation as was done in~\cite{Alonso-Alvarez:2019ssa}} (cf., e.g.~ \cite{PhysRevA.7.368}).  We will then find the number of photons and axions produced through the process $\gamma'\rightarrow \phi +\gamma$, both in the case of (Bose-enhanced) spontaneous decay and in the presence of a photon background (e.g. the CMB) that can help trigger the process. In particular we will show that in certain regions of parameter space the number density of produced photons features parametric enhancement\footnote{See~\cite{Traschen:1990sw,Kofman:1994rk,Shtanov:1994ce,Boyanovsky:1996sq,Kofman:1997yn,Berges:2002cz} for original discussions of parametric resonance, mostly in the context of (p)reheating.}. Finally, we will apply the results to the decay of the dark matter in the early universe and to the CMB spectrum (cf. Sect.~\ref{sec:CMB}) to get the allowed parameter space for $\Gg$\footnote{Besides finding the allowed parameter space for the dark matter, the (stimulated) decay of light dark matter has also been exploited to envisage new detection proposals, for instance in \cite{Arza:2019nta,Caputo:2018vmy}.}.

A detailed analysis of the solution of the system of equations can be found in  Appendices~\ref{app:eoms} and \ref{app:stability}, so here we only highlight the results.  Our starting point is to  consider ${\mathbf{A}}$ and $\phi$ as quantum fields,
with energies given by $\omega(k)=|{\bf{k}}|=k$, and $\o_\phi=\sqrt{k_\phi^2+\mphi^2}$, respectively. Therefore, the energy and momentum conservation of the process $\gamma'\rightarrow \gamma +\phi$ (both stimulated and spontaneous) can be written as
\begin{eqnarray}
\epsilon(k)=\mgamma'-\omega-\omega_\f&\approx& 0\\
{\bf k}_\phi+{\bf k}&\approx& 0
\end{eqnarray}
where the process is at resonance when $\epsilon \rightarrow 0$. In the case of the stimulated decay, the momentum $k$ is the same as the incoming photon. It can be readily seen that the process can enter resonance when
\be 
k=\frac{\mgamma^2-\mphi^2}{2\mgamma},
\ee
thus, in order for it to happen, the hidden photon has to be heavier than the axion, as expected. 

We will consider the initial state as containing a phase space distribution $f_{\gamma,\bf{k}}(0)$ for photons and $f_{\phi,\bf{k}}(0)$ for axions, thus, the number density of photons produced by the decay process is found to be {(see~Appendix~\ref{app:stability})}
\begin{equation}
n_{\gamma}(t)=\int\frac{d^3k}{(2\pi)^3}\left(f_{\gamma,\bf k}(0)\left(\cosh(s_{\bf k}t)^2+\frac{\epsilon_k^2}{4s_{\bf k}^2}\sinh(s_{\bf k}t)^2\right)+f_{\phi,-\bf k}(0)\frac{\OO^2}{s_{\bf k}^2}\sinh(s_{\bf k}t)^2+\frac{\OO^2}{s_{\bf k}^2}\sinh(s_{\bf k}t)^2\right). \label{dngamma}
\end{equation}
Here, we have defined, $\eta \equiv \dfrac{\Gg 
E_0'}4$,  $\OO=\eta\sin\theta\sqrt{\frac{k}{\omega_\phi}}$ and $\sk=\sqrt{\OO^2-\e(k)^2/4}$ and we recall that $\theta$ is the angle between the photon propagation and that of the hidden photon background polarisation.
We note that as long as $\sk$ remains a positive real number, the amount of photons and axions produced in the process is parametrically amplified, i.e. it grows exponentially (see~\cite{Alonso-Alvarez:2019ssa,Yoshida:2017ehj,Hertzberg:2018zte,Arza:2018dcy,Masaki:2019ggg,Carenza:2019vzg,Wang:2020zur,Arza:2020eik,Levkov:2020txo} for detailed discussions in the regular ALP dark matter case). For this to be the case, the condition $-2\OO<\e(k)<2\OO$ has to be fulfilled.

The first and second terms quantify photon emission due to stimulated hidden photon decay triggered by {an} initial occupancy of photons and axions, respectively, while the third term corresponds to photons coming from the spontaneous Bose-enhanced decay. 

We now put all the previous machinery to work. We look into the spontaneous decay of the condensate and, as an important source of stimulated decay, we will consider the thermal photon background from the early universe.

\subsection{Spontaneous decay}
If $m_{\gamma^{\prime}}> m_{\phi}$ the relevant decay process is $\gamma'\rightarrow \phi+\gamma$. The corresponding spontaneous decay rate is given by,
\be
\tau=\frac{1}{\Gamma_{\gamma'\rightarrow \phi\gamma}}=\frac{96\pi}{\Gg^2 \mgamma^3}\left(1-\frac{\mphi^2}{\mgamma^2}\right)^{-3}\approx 2\times 10^{17}\,{\rm{s}}\, \left(\frac{\Gg}{10^{-6}\,{\mbox{GeV}}^{-1}}\right)^{-2}\left(\frac{\mgamma}{1\,{\rm{eV}}}\right)^{-3}.
\ee
In the last step we have assumed $\mgamma/\mphi\ll1$. 
In Fig.~\ref{fig:overview_massless} we show the corresponding constraint for the case {$m_{\phi}=0$ as the orange region.}

\subsection{Stimulated and enhanced spontaneous decay}
At lower masses a stronger constraint can be obtained if it is asked that the Bose condensate  does not enter a regime of parametrically enhanced spontaneous or stimulated decay, as analyzed in detail in~\cite{Alonso-Alvarez:2019ssa} (see~\cite{Abbott:1982af,Preskill:1982cy} for the discussion in the context of the original QCD axion).

While we mostly consider a thermal background as a source of photons for stimulation, our treatment makes explicit the quantum $1/2$ contribution to the occupation number. Therefore, we automatically include also the part of the resonant growth that arises from the vacuum fluctuations. 

Let us start by writing down Eq.~(\ref{dngamma}) in terms of the thermal phase space distribution
\begin{equation}
f_k\equiv f_{\gamma,\bf{k}}(0)=\frac{1}{e^{k/ T}-1}.
\label{CMBpsd}
\end{equation}
Since we are interested only in the parametric resonance regime, where $\sk t\gg 1$, we can approximate
\begin{equation}
n_{\gamma}(t)\simeq\frac{1}{2}\int\frac{d^3k}{(2\pi)^3}\frac{\OO^2}{ \sk^2}\left(f_k+\frac{1}2\right) \,e^{2\sk t}. \label{dsngamma2}
\end{equation}
As already mentioned the extra $1/2$ accounts for the vacuum contribution.

In order to gain more insight into the final result, we will assume $\mphi\ll \mgamma$, thus simplifying the expressions $\OO\simeq \eta\sin\theta$ and $\epsilon(k)\simeq \mgamma-2k$. Since we are summing over all possible directions of propagation of the photons, we will assume, without loss of generality, that $\hat \varepsilon_{dm}$ points in the $\hat z$ direction.
Thus, Eq.~\eqref{dsngamma2} can be written as
\be
n_{\gamma}(t) = \frac{1}{8\pi^2}\int\int dk\,d\theta\, k^2\sin^3\theta\, \eta^2 \left(f_k+\frac{1}{2}\right)\frac{e^{2t\sqrt{\eta^2\sin^2\theta-\epsilon^2/4}}}{\eta^2\sin^2\theta-\epsilon^2/4} .
\ee
To obtain a simple analytical expression we use a saddle point approximation focusing on the dominant growth,
\be
n_\g(t)=\frac{m_{\gamma'}^2\eta}{16\pi}\frac{e^{2\eta t}}{2\eta t}\left(f_{\mgamma/2}+\frac{1}2\right). \label{dsngamma3}
\ee
Here, $f_{\mgamma/2}$ denotes the initial occupation number of photons at energy $k\simeq \mgamma/2$. Moreover, the dependence on the DM polarisation drops out due to the isotropy of the thermal background.

Following the discussion in~\cite{Alonso-Alvarez:2019ssa} (see also~\cite{Abbott:1982af,Preskill:1982cy}) we can now include the effects of expansion. As the universe expands, a photon of frequency $\o= k$, satisfying the resonance condition will be {red-shifted, moving it out of the parametric resonance window. At this point that photon will stop stimulating the decay. After a} time $\d t$, the variation in frequency is $\delta k=m_{\gamma'}H\d t/2$, where $H$ is the Hubble parameter. On the other hand, to be in resonance, the maximum value that $\d k$ can take is $2\eta$, otherwise we would be out of the parametric resonance window. So, in order to obtain the total growth Eq.~(\ref{dsngamma3}) must be evaluated at the time
\be
\d t=\frac{4\eta}{m_{\gamma'}H}. \label{deltat1}
\ee

\begin{figure}[t!]
\centering
    \includegraphics[scale=0.6]{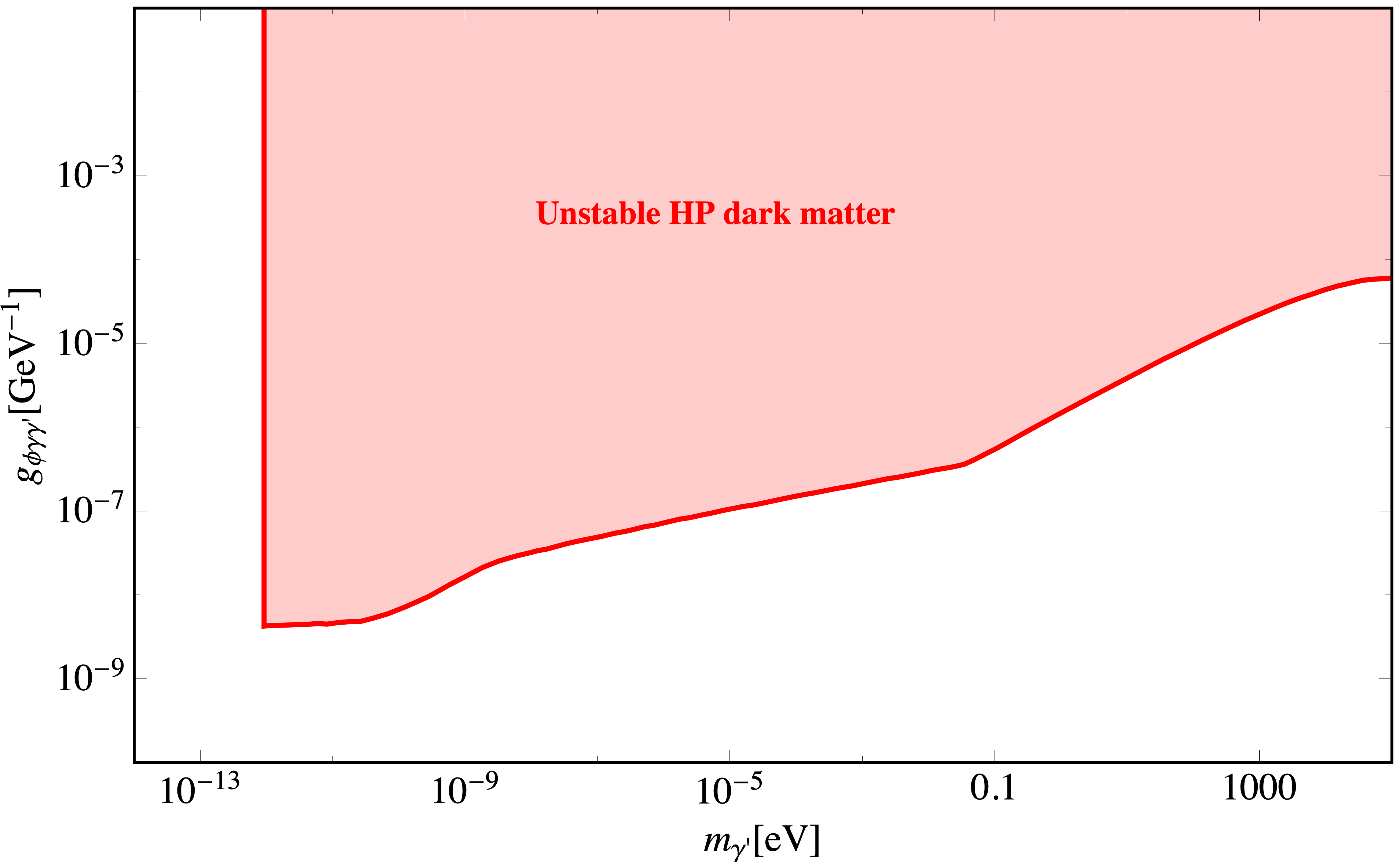}
    \caption{Region where the HP decays to photons and (massless) ALPs due to a parametric resonance, including the effects of the photon plasma mass (we use the values obtained in~\cite{Alonso-Alvarez:2019ssa} which follows the calculations of~\cite{raffelt1996stars,Braaten:1993jw,Redondo:2008ec,Mirizzi:2009iz,Dvorkin:2019zdi}). We also show only the region where the unavoidable loss of coherence is small (see Appendix \ref{sec:coherence}).}
\label{unstable}
\end{figure}
The amplitude of the HP field is related to the dark matter energy density $\rho_{dm}$ by $E_0'=\sqrt{2\rho_{dm}}$. With this the exponent in \eqref{dsngamma3} is $2\eta \d t=\Gg^2\rho_{{dm}}/(m_{\gamma'}H)$. When the universe is dominated by radiation this exponent decreases as $\sim1/\sqrt{t}$ while for matter domination it drops like $\sim1/t$, therefore the parametric resonance is more effective at earlier times. We may also use the approximation $f_{m_{\gamma'}/2}\simeq2T/m_{\gamma'}$. The stability condition is that the photon energy density $\rho_{\gamma}=m_{\gamma'}n_{\gamma'}/2$ must remain smaller than $\rho_{dm}$. {This} leads to
\begin{equation}
\Gg^2<\frac{m_{\gamma'}H}{\rho_{dm}}\ln\left(\frac{64\pi\rho_{dm}}{\sqrt{2Hm_{\gamma'}^5}T}\right). \label{dsgcond1}
\end{equation}
This condition must be fulfilled at every cosmological epoch. Nevertheless, since the exponentials are more effective at earlier times, earlier epochs provide the strongest constraints. That said, we have to be careful that $m_{\gamma'}$ remains bigger than the plasma mass $m_{\gamma}$~\cite{Alonso-Alvarez:2019ssa,Abbott:1982af,Preskill:1982cy}. Fig.~\ref{unstable} shows the parameter space where the HP dark matter is unstable. For each hidden photon mass, condition \eqref{dsgcond1} is evaluated at a cosmological time just after $m_{\gamma}=m_{\gamma'}$.

To conclude this section let us discuss the issue of coherence and comment on its effects. A condition for the validity of the constant field approximation is that all the photons produced by the parametric resonance are coherent over the space-time region when the resonance is active. 
Expressed in momentum space this leads us to require that the width of the resonance is greater than the momentum spread of the hidden photons,
\begin{equation}
    \Delta k_{\rm res}\gtrsim \Delta k_{\rm coh}.
\end{equation}
Using that $\Delta k_{\rm res}\sim \eta$ and the expression, Eq.~\eqref{eq:coherencelinear}, for the coherence scale in the early Universe we have,
\begin{equation}
    \eta\gtrsim\sqrt{H m_{\gamma'}\delta}.
\end{equation}
This can now be compared to the requirement that the red-shift does not destroy the coherence and allows for sufficient growth, Eq.~\eqref{dsgcond1},
\begin{equation}
    \eta\gtrsim \sqrt{H m_{\gamma'}}\left[\ln\left(\frac{64\pi\rho_{dm}}{\sqrt{2Hm_{\gamma'}^5}T}\right)\right]^{1/2}\gtrsim \sqrt{H m_{\gamma'}}.
\end{equation}
The latter is the stronger condition as long as $\delta\lesssim 1$.
As discussed in Appendix~\ref{sec:coherence} this is fulfilled for red-shifts $z\gtrsim 75$. Restricting us to this region of linear structure formation reduces the excluded region to slightly larger masses (cf. Fig.~\ref{unstable}).
A comparison to other constraints is shown in Fig.~\ref{fig:overview_massless}.

%%%%%%%%%%%%%%%%%%%%%%%%%%%%%

\section{Bounds from CMB distortion}{\label{sec:CMB}}
In this section we aim to put further constraints on $\Gg$ by looking into possible distortions of the Cosmic Microwave Background due to processes like (1) $\gamma'\rightarrow \phi+\gamma$ (stimulated HP decay), for masses $\mgamma'>\mphi$, and from (2) $\gamma'+\gamma\rightarrow \phi$ (photon-HP annihilation) in the opposite case $\mphi>\mgamma'$.

Using a standard FRW metric $g_{\mu\nu}={\mathrm{diag}}(1,-R^2,-R^2,-R^2)$ (for more details see Appendix~\ref{FRW}) the photon and axion frequencies in an expanding universe are given by $\omega_k=k/R$ and $\omega_{\phi,k}=\sqrt{k^2/R^2+m_\phi^2}$, respectively, where $R$ is the scale factor and $k$ a comoving wave number. As CMB photons redshift they might eventually match the resonance conditions $\omega_k=(m_{\gamma'}^2-m_{\phi}^2)/(2m_{\gamma'})$ for (1) or $\omega_k=(m_{\phi}^2-m_{\gamma'}^2)/(2m_{\gamma'})$ for (2). Let's choose $t_*$ as the instant when either of them is satisfied for a single frequency and $t_0$ the cosmological time today. Defining $\omega_*=\omega_k(t_*)$ and using that in the matter dominated era the scale factor behaves as, $R\sim t^{2/3}$, the single frequency today that was affected at time $t_*$ is
\be
\omega=\omega_*\left(\frac{t_*}{t_0}\right)^{2/3} \label{ecomega1}
\ee

Before decoupling, CMB photons thermalize rather quickly. In order to be able to observe a distortion in the CMB spectrum today we therefore
assume that the effect takes place between decoupling ($t_d$) and today. Using this, the range of the CMB spectrum that is modified is given by
\be
\omega_*\left(\frac{t_d}{t_0}\right)^{2/3}<\omega<\omega_*. \label{omegarank}
\ee
In this range, the spectral photon energy density $\rho_{\gamma,\bf{k}}$ is distorted by an amount $\delta\rho_{\gamma,\bf{k}}$ which is positive for process (1) and negative for process (2). Assuming that CMB distortions are small, perturbation theory {should} be enough. At {lowest} order, for each comoving wavenumber ${\bf k}$, the space is filled by the CMB photon vector field
\be
{\bf A_{\bf k}}(t)=\hat\varepsilon_{\bf k}A_{\bf k}(t)=\hat\varepsilon_{\bf k}A_{{\bf k},*}\int_{t_*}^tdt'\cos(\omega_kt') \label{cdA1}
\ee
and the hidden photon background field
\be
{\bf E}_{dm}'(t)=\hat\varepsilon_{dm}E_{0,*}'\left(\frac{R(t_*)}{R(t)}\right)^{1/2}\cos\left(m_{\gamma'}(t-t_*)\right). \label{cdEdm1}
\ee
Here ${\bf A_{\bf k,*}}={\bf A_{\bf k}}(t_*)$ and $E_{0,*}'=E_{0}'(t_*)$. At first order, they play the role of sources for the axion field $\phi_{\bf{k}}$. The spectral energy density for the axion field at arbitrary time $t$ is calculated in Appendix~\ref{FRW}. It is given by
\be
\rho_{\phi,\bf{k}}(t)\sim\frac{1}{6}\Gg^2\omega_{\bf k}(t)^2\omega_{\phi,{\bf k}}(t)
\frac{|I_{\bf k}(t)|^2}{R(t)}, \label{cdrhophi1}
\ee
where
\be
I_{\bf k}(t)
=\int_{t_i}^tdt'
\frac{E_{0}'(t')A_{\bf k}(t')}{\sqrt{R(t')^3\omega_{\phi,k}(t')}}e^{-i\int_{t_*}^{t'}dt''\omega_{\phi,k}(t'')}. \label{cdI}
\ee
The number density of axions is the same as the number density of new photons from process (1) and also the same as the number density of eliminated photons in process (2). The correction of the photon energy density therefore is,
\be
\delta\rho_{\gamma,{\bf k}}=\pm\frac{\omega_k}{\omega_{\phi,{\bf k}}}\rho_{\phi,{\bf k}} \label{edeltarhok1}
\ee
where ``$+$" stands for hidden photon decay and ``$-$" for photon-hidden photon annihilation. 
The redshift causes the duration of the resonance to be much smaller than cosmological times. {Therefore we can use,}
\be
R(t)\simeq R(t_*)+\dot R(t_*)(t-t_*)=R_*\left(1+H_*(t-t_*)\right). \label{ecRexp}
\ee
We also have
\be
\omega_k\simeq\omega_*\left(1-H_*(t-t_*)\right) \label{ecomegakexp}
\ee
and
\be
\omega_{\phi,k}\simeq\omega_{\phi,*}-\frac{\omega_*^2}{\omega_{\phi,*}}H_*(t-t_*) \label{ecomegaphikexp}
\ee
where $\omega_{\phi,*}=\omega_{\phi,k}(t_*)$. In this approximation, $E_\text{dm}'$ and $A_{\bf k}$ can be appropriately written as
\bea
E_\text{dm}'(t) &\simeq& E_{0,*}'\cos\left(m_{\gamma'}(t-t_*)\right)  \label{ecsolEdm1}
\\
A_{\bf k}(t) &\simeq& A_{\bf k,*}\cos\left(\omega_*(t-t_*)-\omega_*H_*(t-t_*)^2/2\right). \label{ecsolAk01}
\eea
Plugging (\ref{ecsolEdm1}) and (\ref{ecsolAk01}) into (\ref{cdI}) and neglecting small contributions we get 
\bea
I_{\bf k}^{\pm} &\simeq& \frac{E_{0,*}'A_{{\bf k},*}}{4\sqrt{ R_*^3\omega_{\phi,*}}}\int_{-\infty}^{\infty}d\xi\,e^{\pm im_{\gamma'}H_*\frac{\omega_*}{\omega_{\phi,*}}\xi^2/2} \nonumber
\\
&\simeq& \frac{E_{0,*}'A_{{\bf k},*}}{4\sqrt{ R_*^3\omega_{\phi,*}}}\sqrt{\pm\frac{2\pi}{ im_{\gamma'}H_*}\frac{\omega_{\phi,*}}{\omega_*}}.  \label{ecsolpsok1}
\eea
Notice that this result holds for times after any of the resonances occurred. Thus, the spectral axion energy density just after the resonances is
\be
\rho_{\phi,\bf k}=\frac{\pi}{48}\frac{\Gg^2\omega_*\omega_{\phi,*}}{ R_*^4m_{\gamma'}H_*}E_{0,*}'^2\,A_{\bf k,*}^2. \label{ecsolrhophi1}
\ee
The dark matter amplitude is related with the dark matter energy density 
$\rho_\text{dm}$ by
\be
E_{0}'^2\sim2R^2\rho_\text{dm} \label{ecEsquare1}
\ee
while $A_{\bf k}^2$ is related with the CMB energy density per co-moving wave vector by
\be
A_{\bf k}^2\sim\frac{2R^2}{\omega_k^2}\rho_{\bf k}. \label{ecAsquare1}
\ee
The correction (\ref{edeltarhok1}) in the photon spectral density is simply 
\be
\delta\rho_{\phi,\bf k}=\pm\frac{\pi}{12}
\frac{\Gg^2\rho_{\text{dm},*}}{m_{\gamma'}H_*}\rho_{\bf k}. \label{ecsolrhophi2}
\ee

\begin{figure}[t!]
\centering
    \includegraphics[scale=0.6]{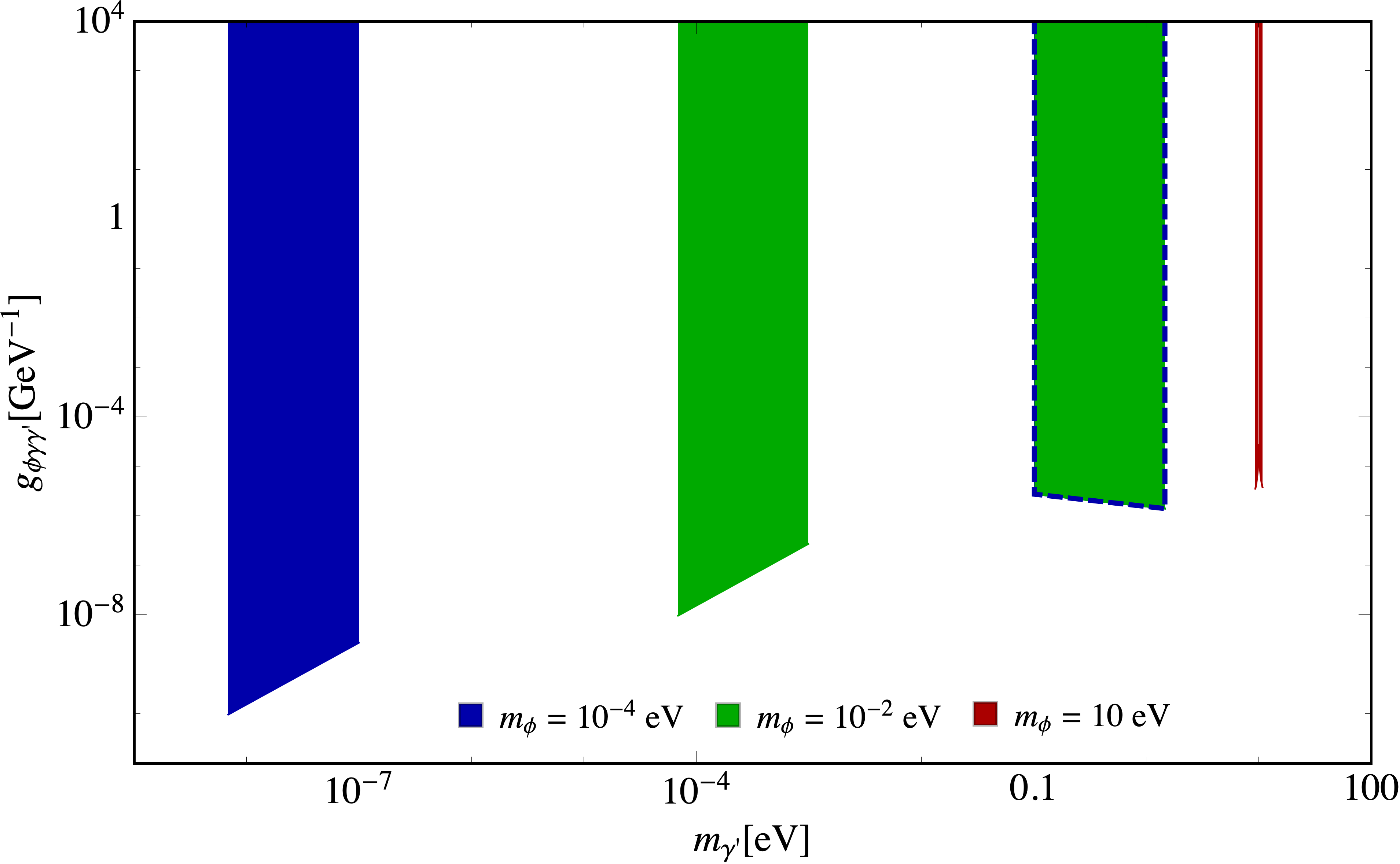}
    \caption{{Limits on the HP-ALP-photon coupling arising from avoiding measurable distortions to the CMB. The two separate regions (for $m_{\phi}=10^{-4}\,{\rm eV}$ as well as $m_{\phi}=10^{-2}\,{\rm eV}$) correspond to the CMB photon induced evaporation into ALPs ($\gamma'+\gamma\to \phi$) (left part) and the CMB photon stimulated decay $\gamma'\to \gamma+\phi$ (right part). In the first case we have a noticeable depletion of CMB photons, in the second we have an increase. We have evaluated the constraints only in the region $z\gtrsim 75$ where the coherence loss from structure formation can be neglected.} }
\label{CMBdist}
\end{figure}

Now, taking into account the facts that $\rho_{\text{dm},*}=\rho_{\text{dm},0}(t_0/t_*)^2$ and $H_*=2/(3t_*)$, the distortions to the CMB spectrum today can be written as
\be
\delta_\omega=\frac{\delta\rho_{\bf k,0}}{\rho_{\bf k,0}}=\pm\frac{\pi}{8}\frac{\Gg^2\rho_{\text{dm},0}\,t_0}{ m_{\gamma'}\chi_\omega^{3/2}} \label{ecdelta1}
\ee
where we have defined $\chi_\omega=\omega/\omega_*$. The accuracy of FIRAS~\cite{Fixsen:1996nj} measurements are of the order of $\delta_\omega\sim10^{-4}$. We can estimate constraints of our model doing $\delta_\omega<10^{-4}$. Using $\rho_{\text{dm},0}\sim1\text{keV/cm}^3$ and $t_0\sim14\times10^9\text{yr}$ we find
\be
\Gg<7.04\times10^{-7}\text{GeV}^{-1}\chi_\omega^{3/4}\left(\frac{m_{\gamma'}}{10^{-5}\text{eV}}\right)^{1/2} \label{ecgconstraint1}
\ee
where $\chi_\omega$ ranges between $(t_d/t_0)^{2/3}\sim10^{-3}$ and $1$. The frequency range of the FIRAS measurements is $2.84\times10^{-4}\text{eV}<\omega<2.65\times10^{-3}\text{eV}$ with a peak at $\omega_\text{peak}=6.63\times10^{-4}\text{eV}$. The spectral energy density of the CMB can be written as
\be
\rho_0(x)=\frac{T^3}{\pi^2}\frac{x^3}{e^x-1} \label{cmbsd}
\ee
where $x=\omega/T$, being $T$ the temperature. The peak is found for $x=x_\text{peak}=3+W(-3e^{-3})$, where $W(x)$ is the Lambert $W$ function. We can see that the CMB temperature is determined by the maximum in the spectral energy density.
To estimate constraints we will evaluate Eq. (\ref{ecgconstraint1}) at $\omega=\omega_\text{peak}$. The ranges for the hidden photon mass that are affected are given by (\ref{omegarank}), we notice that for $m_{\gamma'}>m_{\phi}$
\be
\sqrt{\omega_\text{peak}^2+m_{\phi}^2}+\omega_\text{peak}<m_{\gamma'}<\sqrt{\xi^2\omega_\text{peak}^2+m_{\phi}^2}+\xi\omega_\text{peak} \label{msrank}
\ee
where $\xi=(t_0/t_d)^{2/3}$, and
\be
\sqrt{\xi^2\omega_\text{peak}^2+m_{\phi}^2}-\xi\omega_\text{peak}<m_{\gamma'}<\sqrt{\omega_\text{peak}^2+m_{\phi}^2}-\omega_\text{peak} \label{morank}
\ee
for $m_{\gamma'}<m_{\phi}$. Fig.~\ref{CMBdist} shows the excluded parameters space for CMB distorsions. We use formula (\ref{ecgconstraint1}) for three different values of $m_\phi$ ($10^{-4}$, $10^{-2}$ and $10\text{eV}$) and evaluate it in the ranges (\ref{msrank}) and (\ref{morank}).

\bigskip

As before let us conclude this section by discussing the issue of coherence.
As long as structures can be described by linear perturbation theory we have,
\begin{equation}
    \Delta k_{\rm coh}\sim \sqrt{H m_{\gamma'}\delta}.
\end{equation}
Again this has to be compared with the conditions arising from red-shifting out of the resonance condition.
These read,
\begin{align}
\Delta k_{\rm res}&\sim \sqrt{H m_{\gamma'}} &{\rm for} \quad &m_{\gamma'}>m_{\phi} 
\\\nonumber
\Delta k_{\rm res}&\sim \sqrt{H\frac{m^{2}_{\phi}}{m_{\gamma'}}}>\sqrt{H m_{\gamma'}} &{\rm for} \quad  &m_{\gamma'}<m_{\phi} .
\end{align}
As long as $\delta\lesssim 1$ and linear perturbation theory is valid, the effect of coherence is small compared to the width of the resonance. We therefore again limit ourselves to the region with red-shifts $z\gtrsim 75$ where linear perturbation theory can be applied. This is shown in Fig.~\ref{CMBdist} and similarly in Figs.~\ref{fig:overview_massless} and~\ref{fig:overview_10mhp}.

%%%%%%%%%%%%%%%%%%%%%%%%

\section{Stellar bounds}{\label{sec:solar}}
\subsection{Solar luminosity}
Given its proximity, the sun is the best studied and understood of all stars. 
Unlike other stars, of which we can measure, among others, with different degrees of certainty their mass and metallicity, in the sun we can also add information from neutrino fluxes and helioseismology, achieving a well-established Standard Solar Model\cite{Serenelli:2011py}. Due to this deep understanding, any deviation from  new physics is strictly constrained, making it a powerful tool to test new physics. In particular the energy loss argument has been invoked to study novel weakly interacting particles such as axions~\cite{Raffelt:1987yu, Schlattl:1998fz, Gondolo:2008dd, Vinyoles_2015}, neutrinos~\cite{Raffelt:1999gv} and hidden-photon models \cite{An:2013yfc, Redondo:2013lna}. For any normal star, a new energy loss channel will perturb the stellar object, enforcing it to become more compact, luminous and hotter than the unperturbed configuration~\cite{frieman1987axions}. As a consequence, the nuclear reaction rate increases, eventually modifying its lifetime.
A first conservative  estimate for the sun was $L_x < L^{std}$~\cite{raffelt1999particle}. Later, this was refined to~\cite{Gondolo:2008dd} 
\begin{equation}\label{01luminosity}
    L_x <0.1 L^{std}.
\end{equation}  
Here, the exotic luminosity $L_x$, i.e. the total energy emitted per unit time by non-SM processes, is defined by the relation 
\beq
L_{x}= \int_{V_{\rm sun}} dV\, Q_x,
\eeq
where $Q_x$ is the energy loss rate per volume. It is given by an integral over momentum $\mathbf{q}$,
\beq
Q_x=g_{d} \  \int \frac{d^3 \mathbf{q}}{(2\pi)^3} \ \frac{\Gamma_x \ \omega(q)}{e^{\omega(q) /T}-1},\label{eq:solar_Q}
\eeq
where $g_{d}$ is the number of degree of freedom and $\Gamma_x$ is the decay rate of the process under consideration. This is the elementary object to constrain new physics when we invoke the energy loss argument.

\begin{figure}[t]
    \centering
    \includegraphics[scale=0.3]{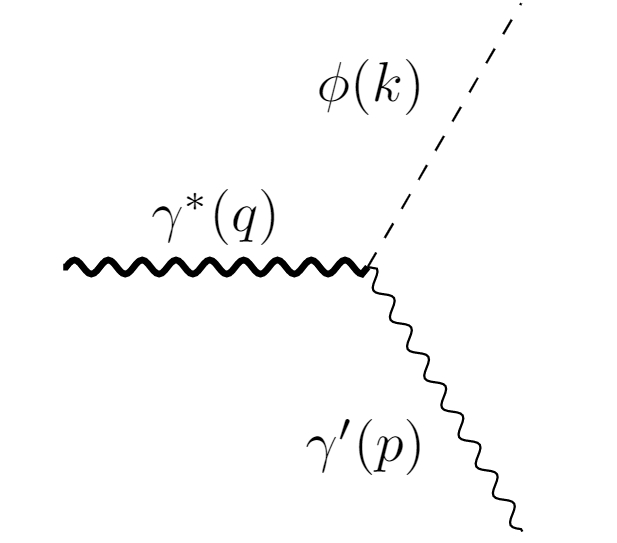}
    \caption{Plasmon decay into a hidden photon and an axion.}
    \label{plasmondecay_feyn}
\end{figure}

For our model, we will focus on the anomalous solar luminosity given by the vertex in the Lagrangian \eqref{lagrangian}. This yields a transverse plasmon decay into a hidden-photon and an axion in a plasma, as shown in Fig.~\ref{plasmondecay_feyn}. This process has already been considered in~\cite{Kohri:2017oqn,Kalashev:2018bra,Gao:2020wer}. We have compared our results\footnote{We are very grateful to Edoardo Vitagliano for noting a mistake in an earlier version of this paper prompting this more detailed comparison.} to those of~\cite{Kalashev:2018bra} and find excellent agreement in the massless limit where our results overlap.

Let us now go through the main steps. If the combined mass of HP and ALP is much smaller than the plasma frequency in the sun, $\o_{pl}$, (massless case), the decay rate is given by\footnote{In Appendix~\ref{appedix_plasmon} we compute in detail the interaction rate.}
\begin{equation}
\Gamma(\gamma^*\rightarrow \g' \phi)=\frac{1}{3}\frac{\Gg^2}{32\pi}\ \frac{\omega_{pl}^4}{\omega},
\end{equation}
this coincides with the result of~\cite{Kalashev:2018bra}.
Here, the plasma frequency is defined as $\omega^2_{pl}=4\pi \alpha n_e / m_e$. $n_e$ is the electronic density and $m_e$ the electron mass. {Inserting this into Eq.~\eqref{eq:solar_Q} we find,}

\beq
Q_{\gamma' \phi}=\frac{\Gg^2}{48\pi^3} \zeta(3)\ \omega_{pl}^4 \ T^3.
\label{eq:Q_massless}
\eeq

The total energy emitted by this process per unit time is obtained by integrating the energy loss rate $Q$ over the solar volume. Using the data from the Standard Solar Model BS05OP of~\cite{Bahcall:2004pz}, we find the anomalous luminosity to be 
\begin{eqnarray}
L_{\gamma' \phi}=g_{10}^2\ 1.84\times 10^{-8}
\ L_{\odot},
\end{eqnarray}
where $g_{10}^{} =\Gg\times 10^{10}\text{GeV}$.
We have already pointed out in Eq.~\eqref{01luminosity} that any exotic energy loss cannot exceed a roughly 10\% fraction of the ordinary solar luminosity. Using this leads to the bound for the coupling
\begin{equation}
    \Gg< 2.3 \times 10^{-7} \ \text{GeV}^{-1}.
\end{equation}
If the mass of the axion and hidden photon are non negligible compared to the plasma mass, ($\omega_{pl}\approx 0.3~$keV in the solar centre), we need to include their effects. The relevant equations are given in Appendix~\ref{appedix_plasmon},  and in particular Eq.~\eqref{finaldecayrate}. By following the same steps outlined for the massless case, we obtain Fig.~\ref{fig:HBSUN_bound}, where we show the result for several axion masses.  As expected, the emission is more efficient for nearly massless particles. As any of the masses (or the sum of the both) gets closer to $\o_{pl}$, the bound gets weaker.

\subsection{Horizontal branch stars}
\begin{figure}
    \centering
    \includegraphics[scale=0.85]{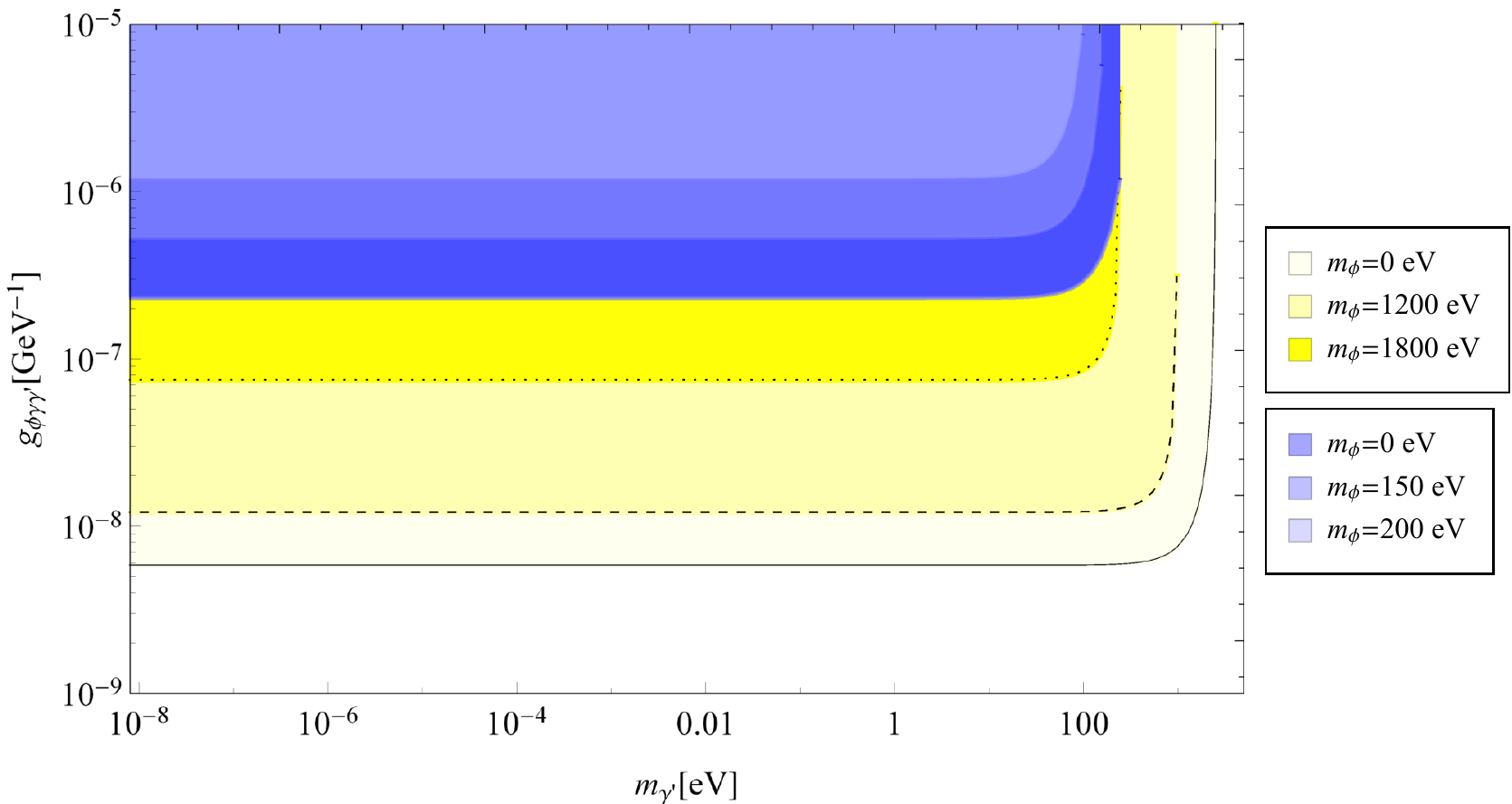}
    \caption{ Bounds on $\Gg$ from solar lifetime (in blue) and from Horizontal Branch stars (yellow) as a function of the dark matter mass, $\mgamma$. The higher density and temperature in the HB stage of a globular cluster provides a much stronger constraint than the solar lifetime. The effect of the axion mass is only noticeable when is  comparable to $\o_{pl}$. For the sun $\omega_{pl}\sim 0.3$~keV, while for the HB the plasma mass is around $\o_{pl}\sim 2$~keV. }
    \label{fig:HBSUN_bound}
\end{figure}

In the previous section we have studied the energy loss in the sun through the new channel opened by the effective vertex $\Gg$, and taken advantage of the knowledge about its structure to impose a stringent bound on the model. However, the sun is cooler and less dense than other astrophysical objects. 
Therefore, other populations such as horizontal branch stars promise improved limits despite the fact that our understanding of them is less developed. The impact of an exotic energy loss channel modifies their inside structure reducing their lifetime. This then changes the time the stars spend in the horizontal branch stage of globular clusters. 

Following the analysis strategy of~\cite{Redondo:2013lna} we approximate
the energy loss rate
via the plasmon decay  $\gamma^* \rightarrow \gamma' + \phi$  as,
\beq \label{energyloss_rate}
\varepsilon_{\gamma' \phi}=\frac{Q_{\gamma' \phi}}{\rho} \sim \Gg^2 \  \frac{\rho \ Y_e^2  \  T^3}{m_u^2 m_e^2} .
\eeq 
Here, $m_u$ is the atomic mass unit, $\rho$ is the density per unit of mass in the star and $Y_e$ is the electron number per baryon.
For simplicity we have also assumed massless axions and hidden photons.

According to the standard model of the sun, at the solar center $\rho\sim153$~g~cm$^{-3}$, $Y_e\sim0.8$ and $T\sim1.3$~keV~\cite{Bahcall:2004pz},  we find $\varepsilon_{\gamma'\phi}\approx 10^{-7}\ \text{erg} \ \text{s}^{-1} \ \text{g}^{-1}$, for $g_{10}=1$.
Compared to the sun, stars in the Horizontal Branch (HB) stage can reach significantly higher densities and temperatures \cite{Raffelt:1987yb}. 
Therefore, the plasmon decay that is possible in our model is more effective. Representative input parameter for this stage are~\cite{raffelt1996stars}: $\rho\sim 10^{4}$~g~cm$^{-3}$, $Y_e\sim 0.5$, $T\sim 10^8$~K, with these values one gets $\varepsilon_{\gamma'\phi}\approx 10^{-1} \ \text{erg} \ \text{s}^{-1} \ \text{g}^{-1}$ for $g_{10}=1$, thus a much tighter constraint on $\Gg$ is expected.  

The compatibility between numerical simulations and statistical observational measurements 
determine that any contribution to the energy loss rate from an additional channel is limited to be~\cite{raffelt1996stars,raffelt1999particle, An:2013yfc}
\beq \label{e_limit}
\varepsilon_{x} < 10 \ \text{erg} \ \text{s}^{-1} \ \text{g}^{-1} \  .
\eeq 

Analogously to the sun's case, we deal with a classical astrophysical plasma because, even if the densities and temperatures increase compared to the solar model, they  are not high enough to change the qualitative physical characteristics. Under these  considerations, we evaluate  Eq.~\eqref{energyloss_rate} for the plasmon decay $\gamma^*\rightarrow \gamma'+\phi$ to get a bound on $\Gg$ from the lifetime of stars on the Horizontal Branch, for massless axions and hidden photons. {Taking the values given above we obtain,}
\beq{\Gg< 6 \times 10^{-9} \ \text{GeV}^{-1}.}
\eeq
As in the solar case, the limit {weakens as the sum of the masses, $\mphi+\mgamma$, approaches} the plasma frequency ($\omega_{pl}\sim 2$~keV in this case). By using the massive decay rate of the process, Eq.~\eqref{eq:int_rate_massive}, we have produced Fig.~\ref{fig:HBSUN_bound} showing the changes in the bound in the case the mass of the particles becomes relevant. As the plasma mass is much higher than in the solar case, the bound {extends to significantly higher masses.} 
 
%%%%%%%%%%%%%%

\section{Laboratory searches and helioscopes}{\label{sec:laboratory}}
The controlled environment of laboratory experiments is very useful in avoiding astrophysical uncertainties and model-dependencies {(cf., e.g.~\cite{Masso:2005ym,Jain:2005nh,Mohapatra:2006pv,Jaeckel:2006xm,Brax:2007ak})}. Also, they could, in principle, provide certain flexibility in the setup, in case new interesting features  emerge. They could then help us to distinguish among different beyond the Standard Model scenarios. 

As a concrete step into testing our model in this section we therefore make use of the latest results of the ALPS~\cite{ehret2010new} light-shining-through-wall (LSW) experiment~\cite{Okun:1982xi,Anselm:1986gz,Anselm:1987vj,VanBibber:1987rq} as well as the helioscopes~\cite{Sikivie:1983ip} CAST~\cite{Anastassopoulos:2017ftl} and IAXO~\cite{Armengaud:2014gea}. We also comment on the effects in optical polarisation experiments.
Thanks to the hidden photon DM background, oscillations of photons into ALPs and vice versa can happen even in vacuum. We therefore also consider the possibility that ALPs, produced in the sun can convert into X-ray photons on their way from the sun.

Before starting on the details let us make a couple of important remarks. As already alluded to, the conversion of ALPs into photons in LSW experiments and helioscopes happens in our scenario only due to the presence of a DM background of HPs\footnote{Indeed the conversion is independent of the employed magnetic field.}. Therefore, in both setups sensitivity to our model is based on the assumption that HPs constitute (all of) the DM and also require that the dark matter survives until today. Vice versa a positive signal would constitute a direct detection of DM.

On the other hand we, again, have to discuss the issue of coherence because the dark matter has a non-vanishing velocity dispersion.
For the pure laboratory experiments (LSW and optical experiments) discussed below the dark matter assumption is needed in the whole process. Coherence is therefore required over the size of the experimental apparatus. For the helioscope-type experiments and observations relying on the production in the sun this assumption is only made for the conversion of the new particles into detectable photons, whereas the production inside the sun is independent of this assumption. Therefore, the coherence requirements only apply for the region in which the conversion into photons takes place.
In any case the experiments are located inside the galaxy. The velocity distribution here is dominated by the effects of structure formation and relatively independent of the initial velocities, therefore the resulting limits are also robust with respect to the production process.
As briefly discussed at the end of Appendix~\ref{sec:coherence} this yields a coherence length,
\begin{equation}
    L_{\rm coh}\sim 10^{3} \frac{1}{m_{\gamma'}}.
\end{equation}
In the following we apply our limits only in the range of HP masses where $L_{\rm coh}\gtrsim L_{\rm experiment}$.

\subsection{Light Shining Through Walls experiments}{\label{subsect:LSW}}
Before going into a proper LSW setup, let us very quickly compute the probability of a photon of energy $\o$, converting into an axion from the process $\gamma+\gamma'\rightarrow \phi$ by using the rotating wave approximation worked out in Appendix~\ref{app:rotatingapprox}. The spirit of this approach is to reduce the system of equations of motion into a first order ones, by keeping only the resonant terms for certain process, whereas the rest gets averaged away. In the case discussed above the resonance of the process occurs when $\Delta_{\g\f}=\mgamma+\o-\o_\f\approx 0$. By keeping only this term in Eqs.~(\ref{ec:partiala}), (\ref{ec:partialphi}) and solving for the photon and axion operators 
we find their evolution in time and consequently over the length of the experiment. 
Accordingly, if we initially have a photon state, the probability that will convert into an axion state is given by
\be
P_{\g \f}(L)=\frac{\OO^2}{\sk^2}\sin^2(\sk L),
\ee
with $\OO=\Gg\sin\theta \sqrt{\frac{\o}{\o_\phi}}$ and $\sk=\sqrt{\OO^2+\Delta_{\g\f}^2/4}$. In the resonant limit, $\sk \rightarrow |\OO|$, and the probability of conversion is maximal, $P_{\g \f}\approx \sin^2(\sk L)$. For small $\sk L$, it is found
\be
P_{\g \f}(L) \approx \frac{\Gg^2  \sin^2\theta\, 
E^{'\,\, 2}_0}{4} \frac{\o}{\sqrt{\o^2+\mphi^2}}L^2.
\ee
This result is essentially the same as in the zero mass limit
of the usual photon-axion oscillations, replacing $g_{\phi\gamma\g}B_{ext}\rightarrow \Gg\sin\theta E'_0$.

\begin{figure}[t]
    \centering
    \includegraphics[width=0.7\textwidth]{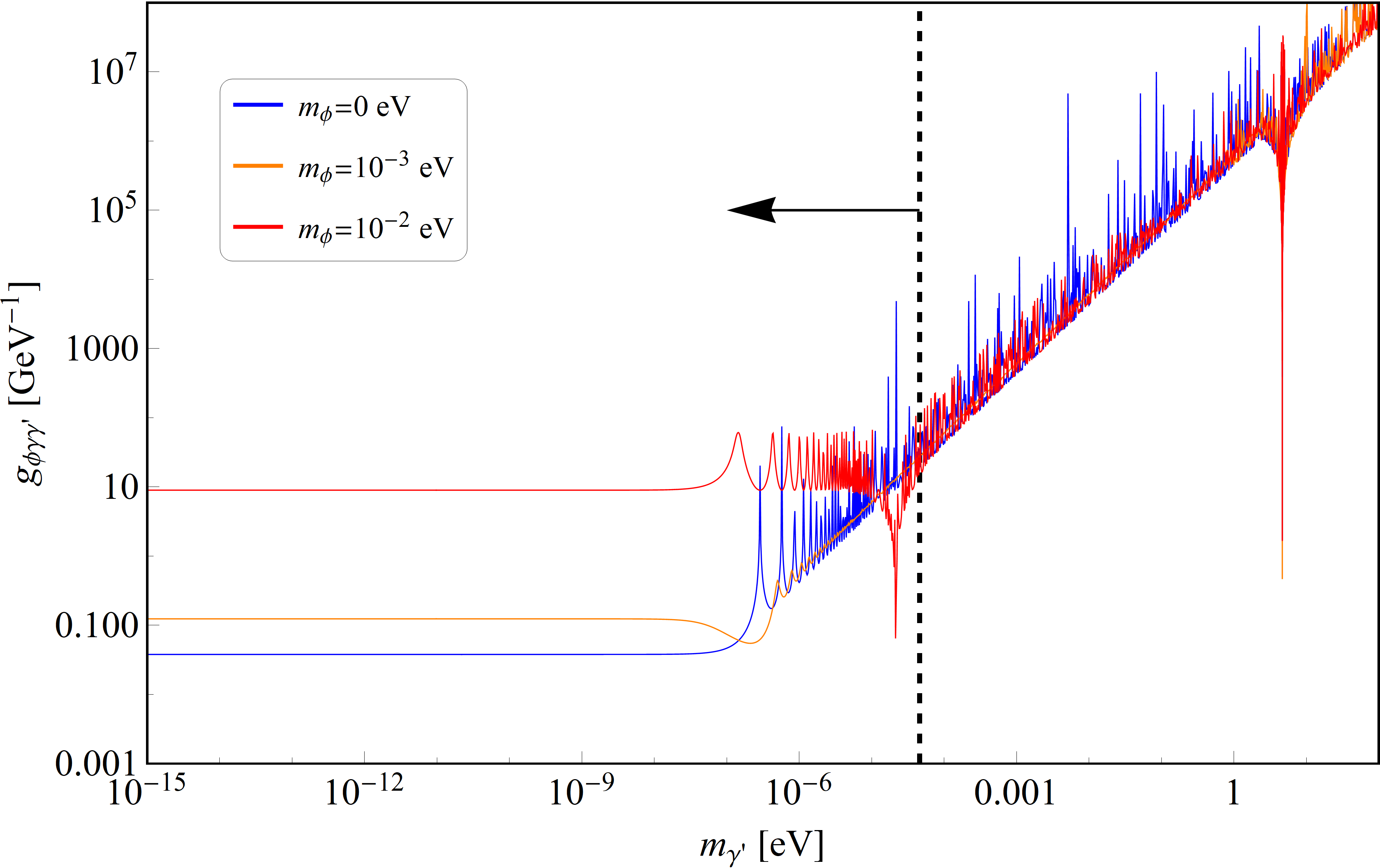}
    \caption{Bound on $\Gg$ as a function of the dark matter mass,  from the light-shining-through walls experiment, ALPS-I~\cite{ehret2010new}. The frequency of the incoming laser is $\o=2.33$~eV, the length of the cavity is $L=4.3$~m and we have use the local dark matter density $\rho_{\rm CDM}=300$~MeV/cm$^3$ and that the DM is randomly oriented in space, {\it i.e.} $\langle \sin^4\theta\rangle$=8/15. We also show the limits where the coherence length is lost due to the momentum dispersion of Hidden Photon DM. The arrow indicates the parameter space where the approximation holds. }
    \label{fig:lsw_bound}
\end{figure}

However, using this approach to compute the oscillation probability in an LSW setup may miss a lot of rich phenomenology, because it could be that more than one kind of process takes place inside the two cavities of the experiment. 
To get a full picture, we have solved the second order equations of motion in a perturbative way, and found the  probability to observe a photon in the regeneration cavity for a LSW setup. The details, together with the full expression for the probability can be {found} in Appendix~\ref{app:LSW}. We have used the data of the experiment ALPS~I~\cite{ehret2010new} to constrain $\Gg$ as a function of the HP mass, as shown in Fig.~\ref{fig:lsw_bound}. 

Similar to~\cite{Arias:2012az} we consider two different alternatives for the polarisation angle $\sin\theta$ appearing in the effective coupling. The first one, is to assume that the DM polarisation is randomly oriented in space. In this case $\langle
\sin^4\theta\rangle=8/15$. Another, more pessimistic scenario, would be to consider the DM points in a particular direction in space, in this case we can obtain a conservative estimate by assuming that the true value $\theta$ is the highest among the 5 $\%$ less likely angles, in that case $\sin^4\theta\sim 10^{-2}$. To produce Fig.~\ref{fig:lsw_bound}, we have considered the DM as randomly oriented in space. We also indicate by an arrow the region in which the coherence condition is fulfilled.

\bigskip
To get a better feeling of the different processes that can occur in the experiment, let us think of it in the following way: when a photon of energy $\o=k$ enters the first generation cavity, there are three different processes that could be possible, depending on the dark matter mass, $\mgamma$ and the energy of the incoming photon, they are:\\

 {\it i) Photon-hidden photon annihilation} $\gamma+\gamma'\rightarrow \f$ : the process is favoured when the  energy conservation relation is fulfilled, $\omega_\f=k+\mgamma$ and the momentum of the ALP produced is $q_+=\sqrt{(k+\mgamma)^2-\mphi^2}$. The momentum transfer is given by $|k-q_+|$ and it is optimized when near zero. 
 For a massless ALP, the momentum transfer is simply $\mgamma$. Thus the probability of photon-hidden photon annihilation is expected to be 
 highest for HP masses close to zero. This is shown as the blue curve in  Fig.~\ref{fig:lsw_bound}. In contrast for a massive ALP, the momentum of the photon can be fully transferred to the ALP, as long as $\mphi>\mgamma$, and if the mass of the hidden photon satisfies 
\be
\mgamma=\frac{\mphi^2}{2k}.
\ee
This enhancement can be seen as the first peak of the red curve of Fig.~\ref{fig:lsw_bound}. On the other side if $m_\phi>\mgamma$, it is not possible to have $|k-q_+|\sim 0$ for small HP masses, and we see that the sensitivity is reduced in this range.

{\it ii) Stimulated photon decay $\gamma\rightarrow \gamma'+\f$ :} This process can be triggered when a photon of energy $k$ such that $k\gg \mgamma, \mphi$, is stimulated to decay into a HP and an ALP by the dark matter background. The process will be favoured when the energy conservation relation is nearly fulfilled, $\omega_\f=k-\mgamma$, and the momentum of the ALP produced is $q_-=\sqrt{(k-\mgamma)^2-\mphi^2}$.
In this case it is impossible to attain {obtain an exact momentum matching between the photon and the ALP. Therefore no resonances are expected. Indeed the missing momentum must be obtained from the uncertainty in the photon momentum, $\Delta k\sim 1/L$, due to its localization inside the cavity of length $L$.
This is most effective when the momentum mismatch is smallest, i.e. for small ALP and HP masses} (much smaller than $\o$). As with the annihilation, for a massless ALP, the momentum transfer is $|k-q_-|\sim \mgamma$. While the momentum transfer for $\mphi>\mgamma$ is $\mphi^2/2\omega$, thus, again this process contributes for the sensitivity of the experiment to get worse when any of the WISP masses increases. 

{\it{iii) Stimulated hidden photon decay $\gamma'\rightarrow \gamma+\phi$ :}}
The process can be realized when an incident photon stimulates the dark matter background to decay. For it to be effective, the energy relation $\o_\f=\mgamma'-k$ has to be fulfilled. Therefore, it can only happen when the mass of the dark matter is bigger than the frequency of the incoming photon (and $\mgamma>\mphi$, of course). The ALP momentum is again $q_-=\sqrt{(\mgamma-k)^2-\mphi^2}$, thus the momentum transfer is $|k-q_-|$, but since $\mgamma>(k, \mphi)$, the process is suppressed for small HP masses. Nonetheless, a resonance can be found, since the momentum transfer can be fully achieved, $|k-q_-|\approx 0$, if $\mphi\neq 0$, such that  
\be
k=\frac{\mgamma^2-\mphi^2}{2\mgamma}.
\ee

\begin{table}[t]
\begin{center}
\begin{tabular}{|l||*{3}{c|}}\hline
\backslashbox{2nd cavity}{1st cavity}
&\makebox[3em]{$A_{\gamma\phi}$}&\makebox[3em]{$SD_{\gamma}$}&\makebox[3em]{$SD_{\gamma'}$}
\\\hline\hline
$A_{\phi\gamma}$ &$\o+2\mgamma$ &$\o$ & $2\mgamma-\o$\\ \hline
$SD_{\phi}$ &$\o$ & $\o-2\mgamma$ &- \\\hline
$SD_{\gamma'}$ &-&-&$\o$\\\hline
\end{tabular}
\caption{Processes that can occur combined in an LSW-type experiment and the resulting energy of the regenerated photon, $\o_{R}$. SD is for stimulated decay, and A for annihilation. }
\label{table:LSW}
\end{center}
\end{table}

From the processes described above, only some of them can contribute simultaneously.
For instance, let us assume there is first an annihilation of the incoming photon of frequency $\o$ together with a DM HP into an ALP (we will denote this process as $A_{\gamma \phi}$), mediated by the dark matter electric field. The resulting ALP has a frequency $\o_\phi=\omega+\mgamma$. In the second cavity, the ALP could be stimulated to decay by the hidden photon background, into a hidden photon and a photon (we will denote this process $SD_{\phi}$). The energy of the regenerated photon is therefore $\o_R=\o$. The momentum transfer of both processes is the same, and therefore, enhanced for small ALP and hidden photon masses (compared to $\o$), with a resonance if $\mphi>\mgamma$ and $\mgamma=\mphi^2/2\o$. An interesting feature of the model is that the regenerated photon can have a different frequency than $\o$. Therefore, a highly tuned detector can miss some of the processes highlighted here.
 
In Tab.~\ref{table:LSW} we show a summary of the processes that are allowed to occur combined in a LSW setup and the resulting energy of the regenerated photon, $\o_R$.

\subsection{Helioscopes}
Having studied the emission of ALPs from the sun in the previous Sect.~\ref{sec:solar} and having obtained the conversion probability of ALPs to photons in LSW in this section we are now ready to combine the results to determine the sensitivity of helioscopes~\cite{Sikivie:1983ip}.

The flux of ALPs from plasmon decay inside the sun and arriving at Earth is given by,
\begin{equation}
\frac{d^{3}N_{\rm ALP}}{dA \,dE\,dt}=\frac{1}{4\pi d^{2}_{\rm Earth}}\int_{V_{\rm sun}} dV\,     \frac{8E^2}{\pi^2}\frac{\Gamma_x }{e^{2E /T}-1}.
\end{equation}
Here, we have used the approximation that each ALP produced in a plasma decay has roughly half the energy of the decaying plasmon ($E=\omega/2$). Moreover, we have neglected the mass of the ALP compared to the energy of the plasmon. In addition, we have also accounted for a factor of two for the two transverse plasmon polarisations. Finally, we note that $\Gamma_{x}$ is the plasmon decay rate in the plasma frame, Eq.~\eqref{finaldecayrate}, i.e. this rate is suppressed by a factor $\omega_{pl}/(2E)$ compared to that in the plasmon rest frame.

This can now be combined with the probability for photon regeneration from an ALP in the HP DM background,
\begin{equation}
    P_{\phi\to\gamma}=\frac{1}{12}\Gg^2{\rho_{\rm CDM}}L^2\left(\Phi(\beta_+L)^2+\Phi(\beta_-L)^2\right),
\end{equation}
where  $\Phi$ is a form factor function defined in Eq.~(\ref{app:eq:form_factor1}) of Appendix~\ref{app:LSW} and $\beta_\pm=E-\sqrt{E^2-m_\phi^2}\pm\mgamma$. Also we have assumed that the DM is randomly oriented in space resulting in $\langle\sin^2\theta\rangle$=2/3.

 \begin{figure}[t]
    \centering
    \includegraphics[width=0.7\textwidth]{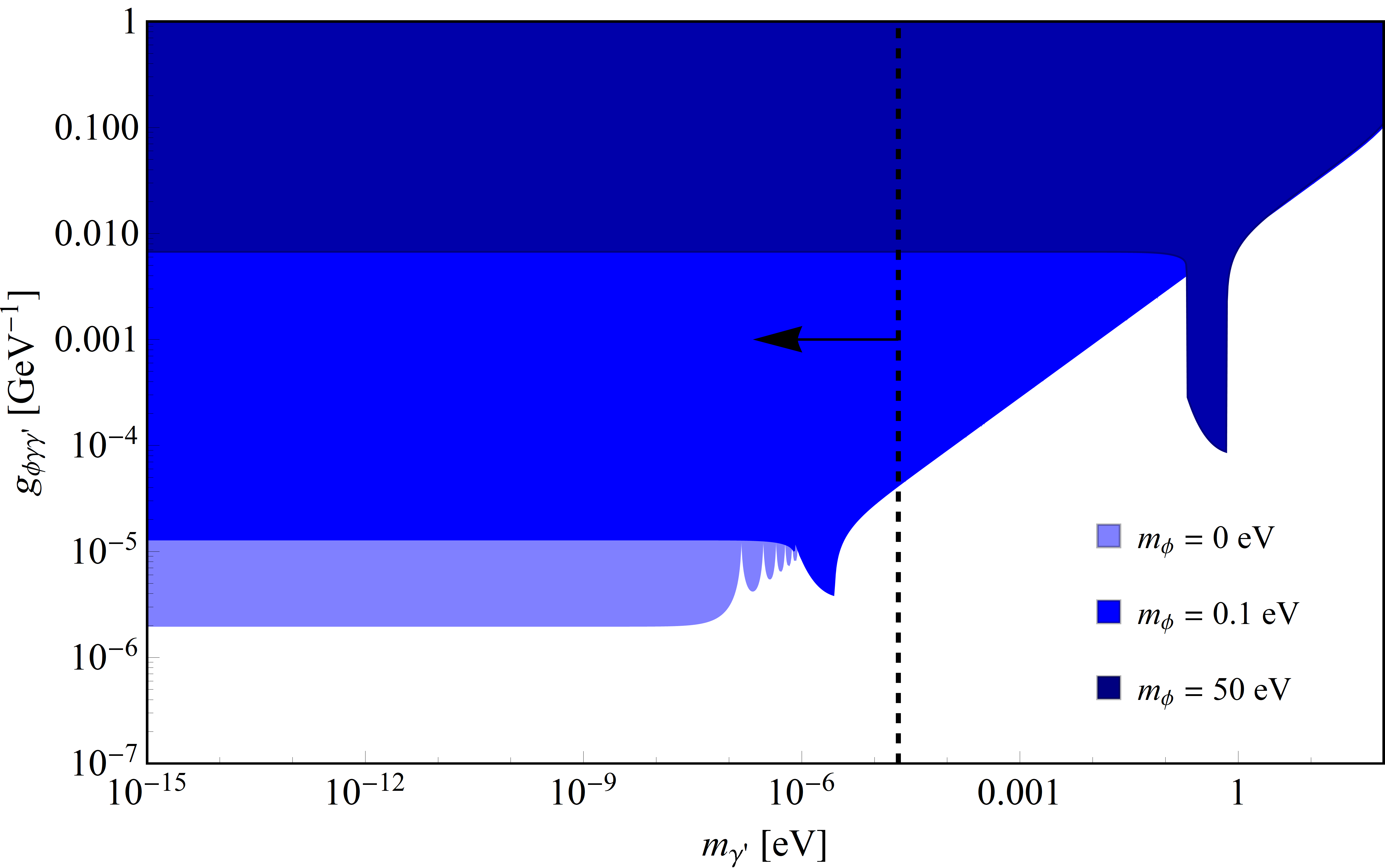}
    \caption{ Bound on $\Gg$ as a function of the dark matter mass, from CAST~\cite{Anastassopoulos:2017ftl}. We use $d_{\rm Earth}=1.5\times10^8\,\text{km}$, $\rho_{\rm CDM}=0.3\,\text{GeV}/\text{cm}^3$ and that the axion-photon conversion takes place in a length $L=9.26\,\text{m}$. The dashed vertical line indicates the region where the DM is coherent over the entire experiment.}
    \label{fig:CAST_bound}
\end{figure}
Fig.~\ref{fig:CAST_bound} gives the excluded parameter space imposed by CAST \cite{Anastassopoulos:2017ftl} for three different ALPs masses; $m_\phi=0$, $1$ and $50\text{eV}$. As before the region where the coherence criterion is satisfied is indicated by an arrow.

The oscillations of the light blue region are a consequence of the form factor $\frac{\sin(m_{\gamma'}L/2)}{m_{\gamma'}L/2}$ in the ALP-photon conversion probability. For massive ALPs this form factor is energy dependent and the oscillations are washed out by the integration over different ALP energies. The bumps of the blue and dark blue regions come from the resonant ALP-photon conversion when the process $\phi+\gamma'\rightarrow\gamma$ is excited. 

In Figs.~\ref{fig:overview_massless} and \ref{fig:overview_10mhp} we show the results for CAST as a red line whereas the red dashed lines give the projection for IAXO~\cite{Armengaud:2014gea}. Let us also emphasize that this bound only holds under the assumption of the existence of a consistent HP dark matter candidate. From Fig.~\ref{fig:overview_massless}, we see that in a big portion of the parameter space where CAST and IAXO are sensitive, the HP is not stable. On the other hand, for $\mphi>\mgamma$, the bound fully applies, see Fig.~\ref{fig:overview_10mhp},  although it is not better than the energy loss constraints.

\subsection[Constraints from solar X-ray observations]{Constraints from solar X-ray observations\footnote{We are deeply indebted to Gonzalo Alonso-\'Alvarez for suggesting that this process could be interesting to look at.\label{foot:thanks}}}
In the helioscope bounds discussed above the field for the conversion of ALPs into detectable photons is provided by the HP DM. No magnetic field is needed. 
Therefore, this conversion can also take place for ALPs on their way from the Sun to Earth. This would lead to an X-ray flux arriving at Earth. This can be measured with suitable rocket flights and satellites see, e.g.~\cite{Sylwester:2012vr,Caspi:2015wca,NOAA}.
However, even without ALPs and HPs the sun is emitting X-rays, e.g. from the solar Corona. This provides a background. As a first conservative estimate we take the observed typical intensity currently measured by the GOES satellite~\cite{NOAA}, of the order of $10^{-8}\,\text{W}/\text{m}^2$ in the energy range $1.57-12.6\,\text{keV}$, as an upper limit on the exotic flux due to ALPs converted into X-ray photons. 
Taking the entire Sun-Earth distance we find, in this range of energy, an intensity
\be
I_\text{sun-earth}\simeq 9.6\times10^{-15}\,\text{W}/\text{m}^2\left(\frac{\Gg}{10^{-10}\,\text{GeV}^{-1}}\right)^4 \label{Ise}
\ee
for HP masses of the order of $10^{-18}\,\text{eV}$ or less. 

The X-ray flux from converted ALPs is expected to be relatively constant in time. We can therefore also use flux observations made during particularly quiet periods of the sun, e.g. those of SphinX~\cite{Sylwester:2012vr}. Taking the energy range of 1.5-3 keV we have a flux background of the order of $7.10\times10^5\,\text{photons}\,\text{m}^{-2}\,\text{s}^{-1}$. In this range of energy the flux provided by ALPs-photon conversion is 
\be
\Phi_\text{sun-earth}\simeq 26.0\,\frac{\text{photons}}{\text{m}^2\times\text{s}}\left(\frac{\Gg}{10^{-10}\,\text{GeV}^{-1}}\right)^4 \label{fluxSphinX}
\ee
To find constraints from SphinX as well as GOES observations, we impose that our signals must be less than twice the corresponding backgrounds (see Fig.~\ref{fig:overview_massless} and Fig.~\ref{fig:overview_10mhp}).

\bigskip

Another option would be to look with a satellite at the dark side of Earth, as suggested for standard axions/ALPs in~\cite{Davoudiasl:2005nh}. Within the energy range $2-10\,\text{keV}$ and for masses smaller than $10^{-12}\,\text{eV}$, solar ALPs lead to an X-ray flux of the order of
\be
\Phi_\text{dark side}\simeq 2.4\times10^{-10}\,\frac{\text{photons}}{\text{m}^2\times\text{s}}\left(\frac{\Gg}{10^{-10}\,\text{GeV}^{-1}}\right)^4\left(\frac{h}{600\,\text{km}}\right)^2 \label{fluxds}
\ee
where $h$ is distance to the detector from the earth surface. In Figs.~\ref{fig:overview_massless} and \ref{fig:overview_10mhp} we show constraints on the parameter space that could be achieved with suitable measurements. As a reference we take (twice) the sensitivity of the RXTE satellite~\cite{Jahoda1996} that was $1.5\times10^{-2}\,\text{photons}\,\,\text{m}^{-2}\,\text{s}^{-1}$ in the range $2-10\,\text{keV}$. We note, however, that this would require taking data with a suitable orientation pointing the satellite towards the Earth but also at the sun. We do not know whether such an aligned measurement has been done and therefore this should be taken only as a sensitivity estimate. 

\subsection{Optical searches: birefringence and dichroism}
Given that only the photon component parallel to $\hat k\times \hat \varepsilon_{dm}$ converts into axions (see Appendix~\ref{app:eoms}), it is expected that an incoming photon beam going through a region of HP-DM will experience a rotation of the initial polarisation plane, and will acquire an extra phase after traveling a distance $L$ from the source. Therefore, experiments looking for birefringence and dichroism, such as PVLAS \cite{DellaValle:2015xxa} can also be sensitive to our model.  In Appendix~\ref{app:LSW} we have computed, from the perturbative analysis of the equations of motion, the corresponding changes in amplitude and phase of an incoming photon beam due to their interaction with the HP background, Eqs.~\eqref{eq:polangle} and \eqref{eq:elip}. 
Constraints from current PVLAS data are weaker than the LSW constraints discussed in subsection \ref{subsect:LSW}.

%%%%%%%%
\section{Conclusions}{\label{sec:conclusions}}
%%%%%%%%%%
In this paper we have entertained the possibility that very light dark matter particles interact not directly but only under the inclusion of an additional light ``messenger'' with the Standard Model (SM). Such a situation arises naturally if the dark matter particle carries a charge under which all SM particles are neutral. While this situation is fairly standard for heavy, WIMP-like dark matter particles it is much less explored for very light (possibly sub-eV) dark matter. The phenomenology of such a dark matter is changed significantly. In particular direct detection becomes much harder.

As a concrete realization of this scenario we have considered a system of a hidden photon (HP) and an axion-like particle (ALP), both carrying an unbroken $Z_{2}$ charge. The dark matter (which we choose to be made from hidden photons) can then only interact under involvement of the axion-like particle.
This leads to qualitative changes compared to the standard case of ALP or HP dark matter. First of all, in the case that the ALP is heavier than the HP, the HP becomes stable\footnote{Pair annihilations into photons are in principle possible but are suppressed by a higher power of the coupling.}, alleviating bounds from decay. 
If the ALP is lighter than the HP (resonant) decays of the dark matter HPs become possible and provide relatively strong constraints (see Fig.~\ref{fig:overview_massless}) although they are not quite as strong as the astrophysical limits and require a sufficiently cold production to ensure coherence.
As long as the mass of the ALP is not too large, astrophysical bounds from the energy loss of stars are applicable and provide the best constraints (cf. Figs.~\ref{fig:overview_massless} and \ref{fig:overview_10mhp}).

Independent of the ALP mass, however, detection in the laboratory is now significantly more challenging. In particular direct detection in experiments used for the detection of ALPs and HPs coupled directly to photons, loose their main detection signal. As an alternative we have looked into the example of light-shining-through-walls (LSW) experiments. In their case, conversion of photons via magnetic fields and back also does not work, because the Standard Model photon can only interact with an ALP and a HP simultaneously. However, in the presence of the HP dark matter background, conversion of a photon into an ALP becomes possible due to the non-vanishing hidden electric field. This process allows both LSW such as ALPS I and II~\cite{ehret2010new,Bahre:2013ywa} and in a similar manner also helisocopes such as CAST and IAXO~\cite{Anastassopoulos:2017ftl,Armengaud:2014gea} to not only become sensitive, but even turn into direct detection experiments. However, in order to overcome the astrophysical and cosmological limits further improvements in sensitivity are needed. At small masses the best sensitivity is obtained from X-ray observations of the sun~\cite{Sylwester:2012vr,Caspi:2015wca,NOAA}. As ALPs produced in the sun can be converted into photons in the hidden photon dark matter background on their way to Earth, this effectively provides a helioscope with gigantic base length. The sensitivity is, however, reduced by the fact that the sun also directly emits X-rays. Reducing or better understanding this background may be a promising way forward.\\

\noindent {\bf Acknowledgements} We would like to thank Gonzalo Alonso-\'Alvarez, Bjoern-Malte Schaefer and Edoardo Vitagliano for very helpful discussions. D.V. is pleased to acknowledge the hospitality of the University of Heidelberg where part of this research was done. P.~A. and D.~V. acknowledge support from FONDECYT project 1161150. P.~A. also thanks AstroCeNT for their hospitality during some stage of this work. A.~A. is happy to acknowledge the University of Santiago of Chile for their hospitality. D.~V.  is supported by VIPO USACH through Beca Convenio Marco. Also D.V. is grateful for support by the \textit{Beca de movilidad} from the Facultad de Ciencia USACH and the \textit{Ministerio de Educación de Chile}.

\appendix
\section{Equations of motion}{\label{app:eoms}}

Our starting point are Eqs.~(\ref{eom_1}) and (\ref{eom_2}):
\bea
\Box {\bf{A}}&=&-\Gg\nabla \phi \times {\bf{E'}}_{dm}, \label{ap:eom1}
 \\
(\Box + m_{\phi}^2)\phi &=& - \Gg {\bf{E'}}_{dm}\cdot{\bf{B}}\label{app:eq:eom2}.
\eea
Let us assume {that} the hidden photon vector is given by ${\bf{E}}'_{dm}=E'_0(t)\,\hat\varepsilon_{dm}$, and that the photon field can be written in terms of 2 transverse polarisations
\be
{\bf{A}}(x,t)=A_+(x,t)\,\hat \varepsilon_++A_-(x,t)\,\hat \varepsilon_-.
\ee
The direction of propagation of the photon field is ${\bf{k}}=k\,\hat k$.
In Fourier space the above equations of motion look like
\bea
\Box_k\,{\bf{A}}(k,t)\cdot (\hat k\times \hat \varepsilon_{dm})&=&-i\Gg\, k\, E'_0(t)\,\phi(k,t)\, \|\hat k\times \hat \varepsilon_{dm}\|^2\\
\left(\Box_k+\mphi^2\right)\phi(k,t)&=&-i\Gg E'_0(t)\, k\,\hat \varepsilon_{dm}\cdot\left(\hat k\times {\bf{A}}(k,t)\right).
\eea
By using the vector identity ${\bf{A}}\cdot\left({\bf{B}}\times {\bf{C}}\right)={\bf{C}}\cdot\left({\bf{A}}\times {\bf{B}}\right)$, we get
\bea
\Box_k{\bf{A}}(k,t)\cdot (\hat k\times \hat \varepsilon_{dm})&=&-i\Gg\, k\, E'_0(t)\,\phi(k,t)\, \|\hat k\times \hat \varepsilon_{dm}\|^2\\
\left(\Box_k+\mphi^2\right)\phi(k,t)&=&i\Gg E'_0(t)\, k\, {\bf{A}}(k,t)\cdot\left(\hat k\times \hat \varepsilon_{dm} \right).
\eea
{Defining} $\mathcal A(k,t)={\bf{A}}(k,t)\cdot \dfrac{(\hat k\times \hat \varepsilon_{dm})}{\|\hat k\times \hat \varepsilon_{dm}\|}$, we finally find 
\bea
\Box_k \mathcal A(k,t)&=&-i\Gg \sin\theta\, k\, E'_0(t)\,\phi(k,t)\label{eq:new_eom1}\\
\left(\Box_k+\mphi^2\right)\phi(k,t)&=&i\Gg \sin\theta\, k\, E'_0(t)\, \mathcal A(k,t).
\label{eq:new_eom2}
\eea
Therefore, only the component of the photon field parallel to the vector $\hat k\times \hat \varepsilon_{dm}$ couples to the hidden photon dark matter background, with a strength $\Gg\, \|\hat k\times \hat \varepsilon_{dm}\|=\Gg\sin\theta$, where $\theta$ is the angle between the direction of propagation of photons and the dark matter polarisation. 
In the case that the dark matter points in random directions in space, it is necessary to take an average of this expression. 

\subsection{Rotating wave approximation}{\label{app:rotatingapprox}}
We start by considering the quantum fields
\bea
\phi({\bf x},t)&=&\int\frac{d^3k}{{(2\pi)^3}} \frac{1}{\sqrt{2\omega_{\phi}(k)}}\left[\phi_{{\bf k}}(t)\,e^{i\left({\bf k}\cdot {\bf x}-\omega_\phi(k)t\right)}+\phi_{{\bf k}}(t)^\dagger\,e^{-i\left({\bf k}\cdot {\bf x}-\omega_\phi(k)t\right)} \right], \label{solphi0}\\
\mathcal A ({\bf x},t)&=&\int\frac{d^3k}{{(2\pi)^3}}\frac{1}{\sqrt{2\omega(k)}}\left[a_{{\bf k}}(t)\,e^{i\left({\bf k}\cdot {\bf x}-\omega(k)t\right)}+a_{{\bf k}}(t)^\dagger\,e^{-i\left({\bf k}\cdot {\bf x}-\omega(k)t\right)} \right], \label{sola0}
\eea
where $\omega_\phi(k)=\sqrt{k^2+\mphi^2}$, $\omega(k)=k$ and $k\equiv |{\bf k}|$.
Moreover, we assume that the amplitudes $a_{\bf k} (t)$ and $\phi_{\bf k} (t)$ are slowly varying functions of time. The creation and annihilation operators satisfy the commutation relations $\left[\phi_{{\bf k}}(t),\phi^\dagger_{{\bf k}'}(t)\right]=\left[a_{{\bf k}}(t),a^\dagger _{{\bf k}'}(t)\right]=(2\pi)^3\delta^3({\bf k}-{\bf k}')$. Plugging (\ref{solphi0}) and (\ref{sola0})  into (\ref{eom_1}) and (\ref{eom_2}),  we can make a further simplification by assuming a rotating wave approximation, {\it i.e.} neglecting the fast oscillating terms.\footnote{This approximation is well known in quantum optics, see for instance \cite{PhysRevA.7.368}.} From there,  we get the coupled system
\bea
\partial_t a_{{\bf k}}&=& {\eta\sin\theta}\sqrt{\frac{k}{\o_\f}}\left( \f_{{\bf k}}e^{-i\Delta_{\g \f}(k)t} + \f_{ {\bf k} }e^{-i\Delta_{\f \g}(k)t} +\f^{\dagger}_{-{\bf k}}e^{-i \e(k) t}   \right), \label{ec:partiala}
\\
\partial_t\phi_{{\bf k}} &=&
- {\eta\sin\theta }\sqrt{\frac{k}{\o_\f}}\left(   a_{{\bf k}}e^{i\Delta_{\gamma\phi}(k)t}+a_{{\bf k}}e^{i\Delta_{\phi\gamma}(k)t}-a_{-{\bf k}}^\dagger e^{-i\epsilon(k)t} \right).\label{ec:partialphi}
\eea
Where the same equations hold for the hermitean conjugate operators, changing ${\bf k}\rightarrow -{\bf k}$. 
Also, we have defined new parameters, such as $\eta \equiv \dfrac{\Gg 
E_0'}4$,  and 
\bea && \Delta_{\g \f}=\o + m_{\g'}-\o_{\f}, \,\,\,\,\, \Delta_ {\f \g}=-\o + m_{\g'}+\o_{\f}, \,\,\,\,\, \epsilon=-\o - \o_{\f}+m_{\g'}.\eea
These coefficients account for energy conservation in different physical processes. For instance, $\Delta_{\g\f}(k)$ and $\Delta_{\f\g}(k)$ account for the energy conservation of the annihilation processes  $\g+\gamma'\rightarrow \f$ and $\f+\gamma'\rightarrow \g$, respectively.  Therefore, we expect that the main contribution to such a process happens when at resonance, i.e. $\Delta_{\g\f}(k)\rightarrow 0$. Indeed, the term $\epsilon(k)$ accounts for the energy conservation of the dark matter decay $\gamma'\rightarrow \gamma+\phi$.
Thus, when a certain process is at or near resonance, we keep only fast oscillating exponentials. From there, the system of equations is quite easy to solve, and will have in general the form
\bea
\partial_t a_{{\bf k}}&=&\OO \phi_{{\bf k}} e^{-i\Delta(k) t}\\
\partial_t \phi_{{\bf k}}&=&\pm\OO a_{{\bf k}} e^{i\Delta(k) t}\label{sol2_rwa},
\eea
where $\OO=\eta\sin\theta \sqrt{\frac{k}{\o_\phi}}$, and $\Delta$ represents the energy conservation relation of the resonant process. The spatial dependence is obtained from $t\rightarrow L$, where $L$ is the length covered on an interval of time $t$. The $\pm$ in front of Eq.~(\ref{sol2_rwa}) accounts for hyperbolic ($+$) or oscillatory solutions ($-$). 

%%%%%%%%%%%%%
\section{Number density of photons and ALPs from DM decay}{\label{app:stability}}
%%%%%%%%%%%%%%%
We want to compute the number of photons produced during the decay process $\gamma'\rightarrow \phi+\gamma$. We start from Eqs.~\eqref{ec:partiala} and \eqref{ec:partialphi}, and assume that we are close to resonance $\e(k)\rightarrow 0$. {We can then ignore all quickly oscillating terms, as they will get averaged away. The relevant system to be solved for the decay process is then    
\bea
\partial_t a_{\mathbf{k}}&=&{\eta\sin\theta}\sqrt{\frac{k}{\o_\f}} \f_{-\mathbf{k}}^{\dagger}e^{-\e(k) t} \equiv \OO\f_{-\mathbf{k}}^{\dagger}e^{-i \e(k) t}
\\
\partial_t \f_{-\mathbf{k}}^{\dagger}&=&{\eta\sin\theta}\sqrt{\frac{k}{\o_\f}} a_{-\mathbf{k}}e^{\e(k) t} \equiv \OO a_{\mathbf{k}}e^{i\e(k) t},
\eea
with the solution
\bea
a_{\mathbf{k}}(t)&=&e^{-i\e(k) t/2}\left[ a_{\mathbf{k}}(0)\left(\cosh(\sk t) +i\frac{\e(k)}{2\sk}\sinh(\sk t) \right)+ \f^{\dagger}_{-\mathbf{k}}(0)\frac{\OO}{\sk}\sinh(\sk t) \right] \label{adecay} \\
\f_{-\mathbf{k}}^{\dagger}(t)&=&e^{i\e(k) t /2}\left[ \f_{-\mathbf{k}}^{\dagger}(0)\left(\cosh(\sk t ) -i\frac{\e(k)}{2\sk}\sinh(\sk t)\right)+a_{\mathbf{k}}(0)\frac{\OO}{\sk} \sinh(\sk t)\right]\label{phidecay} .
\eea
Here $\sk=\sqrt{\OO^2-\e^2(k)/4}$ and $\OO={\eta\sin\theta}\sqrt{\frac{k}{\o_\f}}$.
The  photon and ALP phase space distributions are given by
\bea
f_{\gamma,\bf k}(t) &=& \frac{1}{V}\left<i\right|a_{\bf k}^\dagger(t) a_{\bf k}(t)\left|i\right>
\\
f_{\phi,\bf k}(t) &=& \frac{1}{V}\left<i\right|\phi_{\bf k}^\dagger(t) \phi_{\bf k}(t)\left|i\right>.
\eea
Then, using Eq.~\eqref{adecay}, we find
\be
f_{\gamma,\bf k}(t)=f_{\gamma,\bf k}(0)\left(\cosh(s_{\bf k}t)^2+\frac{\epsilon_k^2}{4s_{\bf k}^2}\sinh(s_{\bf k}t)^2\right)+f_{\phi,-\bf k}(0)\frac{\OO^2}{s_{\bf k}^2}\sinh(s_{\bf k}t)^2+\frac{\OO^2}{s_{\bf k}^2}\sinh(s_{\bf k}t)^2. \label{dfgamma1} 
\ee
Here, $f_{\gamma,\bf k}(0)$ is the initial occupation number.

Integrating over phase space we obtain the photon number density,
\bea
n_{\gamma}(t)\!\!\!&=&\!\!\!\int\frac{d^3k}{(2\pi)^3}f_{\gamma,\bf k}(t)\\
\!\!\!&=&\!\!\! \int\frac{d^3k}{(2\pi)^3}\left(f_{\gamma,\bf k}(0)\left(\cosh(s_{\bf k}t)^2+\frac{\epsilon_k^2}{4s_{\bf k}^2}\sinh(s_{\bf k}t)^2\right)+f_{\phi,-\bf k}(0)\frac{\OO^2}{s_{\bf k}^2}\sinh(s_{\bf k}t)^2+\frac{\OO^2}{s_{\bf k}^2}\sinh(s_{\bf k}t)^2\right). \notag \label{dngamma1}
\eea
We can follow the same steps to compute the ALP number density
\begin{equation}
n_{\phi}(t)=\int\frac{d^3k}{(2\pi)^3}\left(f_{\phi,\bf k}(0)\left(\cosh(s_{\bf k}t)^2+\frac{\epsilon_k^2}{4s_{\bf k}^2}\sinh(s_{\bf k}t)^2\right)+f_{\gamma,-\bf k}(0)\frac{\OO^2}{s_{\bf k}^2}\sinh(s_{\bf k}t)^2+\frac{\OO^2}{s_{\bf k}^2}\sinh(s_{\bf k}t)^2\right). \label{dnphi1}
\end{equation}

\section{Axion production in a FRW metric: First order calculation} \label{FRW}

Here we provide a first order computation for an axion field, produced when the interaction between a photon background ${\bf A}$ and the hidden photon dark matter field ${\bf E}'_{dm}$ takes place in a FRW metric. 

Our choice of metric is,
\begin{equation}
    g_{\mu\nu}={\mathrm{diag}}(1,-R^2,-R^2,-R^2),
\end{equation}
with the scale factor $R(t)$. $H=\dot R/R$ is the Hubble parameter. Moreover we use the convention $A_\mu\equiv\{A_0,-{\bf A}\}$ and we often use the electric field defined as, 
\begin{equation}
    E_{i}=F_{0i}.
\end{equation}
We note that this definition deviates by a factor of $R$ from the naive ``physical'' field $\sqrt{E^{i}E_{i}}$ (with no summation implied).

In the expanding Universe Eqs.~\eqref{eom_1} and \eqref{eom_2} are replaced by
\be
\left(\partial_t^2+3H\partial_t-\frac{\nabla^2}{R^2}+m_\phi^2\right)\phi
=-\Gg\frac{{\bf E}_{dm}'}{R^3}\cdot\nabla\times{\bf A}. \label{FRWeqphi1}
\ee
In Fourier space the above equation can be written as 
\be
\left(\partial_t^2+3H\partial_t+\frac{k^2}{R^2}+m_\phi^2\right)\phi_{\bf k}= igk\sin(\theta)\sin(\varphi_{\bf k})\frac{E_{dm}'}{R^3}A_{\bf k}, \label{FRWeqphik1}
\ee
where $\varphi_{\bf k}$ is the angle formed by the polarisation vectors of ${\bf A}$ and ${\bf E}'_{dm}$.

First we eliminate the term $3H\partial_t$ by defining $\phi_{\bf k}=R^{-3/2}\tilde\phi_{\bf k}$. We get
\be
\left(\partial_t^2+\omega_{\phi,k}^2\right)\tilde\phi_{\bf k}=igk\sin(\theta)\sin(\varphi_{\bf k})R^{-3/2}E_{dm}'A_{\bf k} \label{FRWeqphiktilde11}
\ee
where $\omega_{\phi,k}^2=\omega_k^2+m_\phi^2$ and $\omega_k=k/R$. Looking for resonant solutions, we write the ansatz
\be
\tilde\phi_{\bf k}(t)=\psi_{\bf k}(t)\zeta_k(t) \label{FRWansphiktilde}
\ee
where $\psi_{\bf k}$ is a slowly varying function and $\zeta_k(t)$ satisfies the homogeneous equation
\be
\left(\partial_t^2+\omega_{\phi,k}^2\right)\zeta_k(t)=0 \label{FRWeqzeta}
\ee
whose solution is
\be
\zeta_k(t)=\omega_{\phi,k}(t)^{-1/2}e^{i\int_{t_*}^tdt'\omega_{\phi,k}(t')}. \label{FRWsolzeta}
\ee
Neglecting second derivatives of $\psi_{\bf k}$, Eq.~\eqref{FRWeqphiktilde11} gives the solution
\be
\psi_{\bf k}(t)=\frac{i}{2}gk\sin(\theta)\sin(\varphi_{\bf k})\int_{t_i}^tdt'\frac{E_{0}'(t')A_{\bf k}(t')}{R(t')^{3/2}\dot\zeta_k(t')} \label{FRWsolpsik1}
\ee
where $t_i$ is some initial time when $\psi_{\vec k}$ is zero. Now, the axion energy density 
\be
\rho_{\phi,{\bf k}}\sim2\omega_{\phi,k}^2|\phi_{\bf k}|^2 \label{FRWrhophi1}
\ee
can be computed straightforwardly. Averaging over $\theta$ and $\varphi_{\bf k}$ we find
\be
\rho_{\phi,{\bf k}}(t)\sim\frac{1}{6}g^2\omega_{\bf k}(t)^2\omega_{\phi,{\bf k}}(t)
\frac{|I_{\bf k}(t)|^2}{R(t)}, \label{FRWrhophi2}
\ee
where
\be
I_{\bf k}(t)
=\int_{t_i}^tdt'
\frac{E_{0}'(t')A_{\bf k}(t')}{\sqrt{R(t')^3\omega_{\phi,k}(t')}}e^{-i\int_{t_*}^{t'}dt''\omega_{\phi,k}(t'')}. \label{FRWI1}
\ee

\section{Coherence scale of light DM}\label{sec:coherence}
An estimate for the coherence length $L_{\rm coh}$ is that the phases of two particles that have a typical momentum difference have a phase difference of order 1 over the coherence region,
\begin{equation}
    (k_{1}-k_{2})L_{\rm coh}\sim  \Delta k_{\rm coh}L_{\rm coh}\lesssim 1.
\end{equation}
In other words the coherence length is given by the inverse width in momentum space of the hidden photon distribution.

As we are dealing with non-relativistic particles,
\begin{equation}
   \Delta k_{\rm coh}\sim m_{\gamma'}\Delta v,
\end{equation}
where $\Delta v$ is the width of the velocity distribution.

In principle we can define a coherence time in an analogous way. However, due to the non-relativistic nature of dark matter the coherence length $\sim 1/(m_{\gamma'}\Delta v)$ in natural units is smaller than the coherence time $1/(m_{\gamma'}\Delta v^2)$, where $\Delta v^2$ is the typical velocity squared spread $\sim (\Delta v)^2$.
We therefore focus on the coherence length and the corresponding momentum spread.

\subsection{Coherence scale in the early Universe}
Let us now determine the typical velocity fluctuations in the dark matter\footnote{We are very grateful to Bjoern-Malte Schaefer and Gonzalo Alonso-\'Alvarez for clarifying discussions on this.}. In the early Universe we can take density and velocity fluctuations to be linear and estimate them using cosmological perturbation theory (cf., e.g.,~\cite{Peebles}). The fluctuations, $\Phi$, of the gravitational potential are given by the Poisson equation,
\begin{equation}
    \Delta \Phi=4\pi G (1+3w)\rho\delta \sim H^{2}\delta, 
\end{equation}
where $\delta\sim \delta\rho/\rho$ indicates the typical size of the fluctuation in the energy density\footnote{A somewhat more precise definition would be to use the amplitude of the dimensionless power spectrum $\Delta^{2}(k)=k^3P(k)/(2\pi)^2\sim \delta^2$, evaluated at the time of interest.}. $w$ is the equation of state parameter, $w=1/3$ for radiation and $w=0$ for matter.
We can now estimate the typical velocities from the typical gravitational acceleration $\nabla \Phi$ acting over a Hubble time,
\begin{equation}
    \Delta v\sim \nabla \Phi \frac{1}{H}.
\end{equation}
Using $\Delta k_{\rm coh}=m_{\gamma'}\Delta v$ and evaluating the derivatives in momentum space at the scale $\Delta k_{\rm coh}\sim L_{\rm coh}$ we obtain,
\begin{equation}
\label{eq:deltacoherence}
    \Delta k_{\rm coh}\sim \sqrt{H m_{\gamma'}\delta}.
\end{equation}

At sufficiently early times $\delta\sim 10^{-4}-10^{-5} \lesssim 1$ and our estimate based on linear perturbation theory is applicable.
However, during matter domination fluctuations start to grow and at some point become non-linear. A reasonable estimate for the scale at which this occurs is when the variance of $\delta$ is of order of one (cf., e.g.,~\cite{Bartelmann}),
\begin{equation}
    \int^{k_{NL}}_{0} dk\, \frac{k^2 P_{\rm lin}(k,z)}{2\pi^2}\sim 1.
\end{equation}
Here, $P_{\rm lin}(k,z)$ is the linear power spectrum at the red-shift $z$.
Using the simple fitting formula for today's linear matter power spectrum from~\cite{Bardeen:1985tr,Sugiyama:1994ed}, and the linear growth function from~\cite{Peebles,Linder:2003dr} (cf. also, e.g.,~\cite{Schaefer}) normalized via $\sigma_{8}$ (cf., e.g.~\cite{Sugiyama:1994ed,Schaefer}) to the recent Planck data~\cite{Aghanim:2018eyx} we estimate that for $z\gtrsim 75$ all (physical) scales with $k\lesssim 10^{-11} \,{\rm eV}$ are still linear.

However, we may worry that at even smaller length scales non-linearities set in earlier.
In the case of light bosons this is however, prevented by the fact that small scale fluctuations at a length scale $1/k$ correspond to momenta of size $k$. In turn this entails a non-vanishing velocity $v\sim k/m_{\gamma'}$. This leads to an effective Jeans (momentum) scale (see, e.g.,~\cite{Marsh:2015xka}) above which (linear) structures do not grow and are therefore relatively suppressed. For our case this is given by (cf.~\cite{Marsh:2015xka}),
\begin{equation}
k_{\rm Jeans}\sim \sqrt{m_{\gamma'} H}\sim 10^{-11}\,{\rm eV}\left(\frac{m_{\gamma'}}{10^{5}\,{\rm eV}}\right)^{1/2}\left(\frac{H}{10^{-27}\,{\rm eV}}\right)^{1/2}.
\end{equation}
Now, if the Jeans length scale $1/k_{\rm Jeans}$ is bigger than the scale $1/k_{NL}$ at which the first non-linearities can form, i.e. $k_{\rm Jeans}\lesssim k_{\rm NL}$ we can take all structures to be linear and $\delta\lesssim 1$. 

\subsection{Coherence scale inside the galaxy}
It is clear that inside the galaxy we cannot use linear perturbation theory.
Here, we can however, rely on the standard estimate that the dark matter particles at our position inside the galaxy have typical velocities,
\begin{equation}
    v\sim\Delta v\sim 10^{-3}.
\end{equation}
This gives us the typical coherence scale/length,
\begin{equation}
\label{eq:coherencelinear}
    \Delta k_{\rm coh}\sim \frac{1}{L_{\rm coh}}\sim \frac{1}{m_{\gamma'}\Delta v}\sim 10^{3}\frac{1}{m_{\gamma'}}.
\end{equation}

\subsection{Coherence length in scenarios with non-vanishing initial momentum}
\label{app:productioncoherence}
If the bosonic field is present during inflation, the misalignment mechanism~\cite{Abbott:1982af,Preskill:1982cy,Dine:1982ah,Nelson:2011sf,Arias:2012az} produces extremely cold particles with negligible initial momentum.
Here, we want to briefly discuss this situation where the initial momentum is non-vanishing.

In many scenarios where the bosonic particles, are produced from fluctuations, from misalignment after inflation, or from decays of precursor particles or topological defects, initial velocities are non-vanishing~\cite{Sikivie:2006ni,Arias:2012az,Graham:2015rva,Cosme:2018nly,Alonso-Alvarez:2018tus,Tenkanen:2019aij,AlonsoAlvarez:2019cgw,Ema:2019yrd,Ahmed:2020fhc,Agrawal:2018vin,Co:2018lka,Dror:2018pdh,Bastero-Gil:2018uel,Sikivie:1982qv,Davis:1986xc,Harari:1987us,Davis:1989nj,Dabholkar:1989ju,Hagmann:1990mj,Battye:1993jv,Battye:1994au,Ringwald:2015dsf,Long:2019lwl}). Also self-interactions can cause a fragmentation of initially homogeneous fields and corresponding non-vanishing momenta (cf.~\cite{Berges:2019dgr}).
Typical initial momenta are often of the order of,
\begin{equation}
    k_{1}\sim \kappa H_{1}\sim \kappa m_{\gamma '},
\end{equation}
where the index 1 indicates the time when $H_{1}\sim m_{\gamma'}$ and corresponds to the point when a homogeneous field would start performing weakly damped oscillations. Moreover, $\k$ is a model-dependent numerical factor quantifying the typical momentum scale at production.

Using that during radiation domination,
\begin{eqnarray}
k(R)&\sim& p_{1}\left(\frac{R_{1}}{R}\right)
\\\nonumber
H(R)&\sim& H_{1}\left(\frac{R_{1}}{R}\right)
\end{eqnarray}
we find that,
\begin{equation}
    \Delta k(R)_{\rm coh}\sim \kappa \sqrt{H(R)m_{\gamma'}}.
\end{equation}
This has the same parametric dependence as the momentum, Eq.~\eqref{eq:deltacoherence}, imprinted by linear structure formation. However, $\kappa$ is often of the order of 1 or larger in many cases. Typical values in the models~\cite{Sikivie:2006ni,Arias:2012az,Graham:2015rva,Cosme:2018nly,Alonso-Alvarez:2018tus,Tenkanen:2019aij,AlonsoAlvarez:2019cgw,Ema:2019yrd,Ahmed:2020fhc,Agrawal:2018vin,Co:2018lka,Dror:2018pdh,Bastero-Gil:2018uel,Sikivie:1982qv,Davis:1986xc,Harari:1987us,Davis:1989nj,Dabholkar:1989ju,Hagmann:1990mj,Battye:1993jv,Battye:1994au,Ringwald:2015dsf,Long:2019lwl,Berges:2019dgr} are  
\begin{equation}
    \kappa \sim (1-10^2).
\end{equation}
But larger values are also feasible.
Therefore, this presents a more severe limitation of the coherence length.

After matter-radiation equality the scaling slightly changes. Similarly to the estimate above we obtain,
\begin{equation}
    \Delta k(R)_{\rm coh}\sim \kappa \sqrt{H(R)m_{\gamma'}}\left(\frac{H(R)}{H_{\rm eq}}\right)^{\frac{1}{6}}.
\end{equation}

\bigskip

We note that the values of $\kappa$ are limited by the requirement of successful structure formation which requires that the velocity at equality is not too large.
Indeed, for velocities,
\begin{equation}
    v_{\rm eq}\sim 10^{-3}
\end{equation}
dark matter starts to be warm (cf., e.g.~\cite{Viel:2005qj}) and much larger velocities are excluded. 
Evaluating the velocity at equality $H_{\rm eq}\sim 2\times 10^{-28}\,{\rm eV}$ we have,
\begin{equation}
    v_{\rm eq}=\frac{P(R_{\rm eq})}{m_{\gamma'}}
    \sim \kappa \sqrt{\frac{H_{\rm eq}}{m_{\gamma'}}}\lesssim 10^{-3}
\end{equation}
we find the estimate,
\begin{equation}
    \kappa\lesssim \left(\frac{m_{\gamma'}}{10^{-22}\,{\rm eV}}\right)^{\frac{1}{2}}.
\end{equation}

\section{Plasmon decay rate \texorpdfstring{$\gamma^*\rightarrow \phi+\gamma'$}{gammatophigammaprime}}\label{appedix_plasmon}
The decay rate of the process depicted in Fig.~\ref{plasmondecay_feyn}, $\gamma^*(q)\rightarrow \phi(k)+\gamma'(p)$ is 
\begin{equation} \label{plasmon_decay}
\Gamma_{\gamma^*\rightarrow \phi\gamma'}=\frac{1}{2\omega_{pl}({\bf q})}\int \frac{d^3\mathbf{p}}{(2\pi)^3\ 2\omega_{\gamma'}({\bf p})}\int \frac{d^3\mathbf{k}}{(2\pi)^3\ 2\omega_{\phi}({\bf k})} \ Z_T|\mathcal{M}|^2(2\pi)^4 \delta^{(4)}(k + p-q),
\end{equation}
where $q=(\o, \bf{q})$ is the plasmon 4-momentum  and the dispersion relations for the ALP and hidden photon are $\o_\phi^2=\mphi^2+{\mathbf{k}}^2, \o_{\gamma'}^2=\mgamma^2+{\mathbf{p}}^2$, respectively. For the renormalization factor we use the approximation $Z_{T}\approx 1$~\cite{raffelt1996stars}.

The scattering matrix is given by 
\bea
\mathcal{M}=\Gg e_{\mu}^{*}(p)e_{\nu}(q)\epsilon^{\mu\nu\alpha\beta}p_{\alpha}q_{\beta}.
\eea Since the plasma sets a preferred reference frame in the system, the decay rate  shall be computed in the plasma frame. \footnote{We thank Edoardo Vitagliano for pointing this out to us.} 
However, by ignoring the contribution of the longitudinal plasmon\footnote{In classic stars, the energy loss contribution from longitudinal plasmons is suppressed relative to transverse plasmon due to the smaller phase space. A similar behavior can be found in the case of a non-standard neutrino dipole moment\cite{raffelt1996stars}. } the dispersion relation for the transverse components is $-\omega^2+{\bf q}^2+m_T^2=0$. Either in the  classical limit ($m_e\gg (q, T )$) or in the relativistic limit ($\omega\sim |\bf q|$), the effective transverse mass $m_T$ is given by \cite{raffelt1996stars}
\be
m_T^2=\opl^2\left(1+\frac{1}2 G(v_*|{\bf{ q}}|^2/\o^2)\right),
\ee with $v_*$ the electron velocity and $G$ a function with limiting behavior $G(0)=0$. For a classical plasma the typical velocity of an electron satisfies $v_*\ll 1$. Therefore we can use $G\approx 0$. Then the dispersion relation approaches that of a massive particle with a fixed mass, $\opl$. This has a Lorentz invariant form, and there is no Lorentz violation due to the plasma frame at this level of approximation.

Let us assume without loss of generality that the plasmon travels in the $z-$direction. Our strategy will be to compute the decay rate in the plasmon rest frame, and then boosting the result to the plasma frame, by applying the boost factor $\opl/\o$.}\footnote{We have checked that the result directly in the plasma frame coincides with the derivation in the rest frame and then applying the boost, as shown here.} In the rest frame the 4-momentum vector is $q=(\omega_{pl},0,0,0)$. We will be using the following relations for the polarization vectors of the hidden photons and plasmon, respectively  \cite{greiner2013field}
\bea
\sum_{b=1}^3e_\mu^{(b)}e_{\mu'}^{(b)}&=&\eta_{\mu\mu'}-\frac{p_\mu p_{\mu'}}{\mgamma^2},\\
\sum_{a=1}^3 e_i^{(a)} e_j^{(a)}&=&\delta_{ij}.
\eea
Thus, averaging the contribution from both transverse polarizations, gives
\bea
\overline{|\mathcal{M}|^2} &=&
\frac{1}{2}\,\Gg^2\left(-2 p^2 q^2 + 2(p\cdot q)^2 - \opl^2\bold{p}^2 \sin^2\theta \right),
\eea
where $\theta$ is the angle between the $z-$direction and ${\bf p}$.
Considering the on-shell equation $2p\cdot q=q^2+p^2-k^2$, we finally get
\bea
\overline{|\mathcal{M}|^2} =
\frac{\Gg^2}{16}\left[ \left(\opl^2-\left(\mgamma+\mphi\right)^2\right)\left(\opl^2-\left(\mgamma-\mphi\right)^2\right)\right]\left(1+z^2 \right)\equiv g_{\rm eff}^2(1+z^2),
\eea
with $z=\cos \theta$. Replacing in the decay rate eq.~(\ref{plasmon_decay}) we  find
\begin{equation}
\Gamma_{\gamma^*\rightarrow \phi\gamma'}=\frac{1}{2\opl} g_{\rm eff}^2 \int \frac{d^3\mathbf{p}}{(2\pi)^3\ 2\omega_{\gamma'}}\int \frac{d^3\mathbf{k}}{(2\pi)^3\ 2\omega_{\phi}}\left(1+z^2\right)
(2\pi)\delta(\omega-\omega_{\phi}-\omega_{\gamma'})(2\pi)^{3}\delta^{(3)}(\mathbf{k}+\mathbf{p})
\end{equation}
\begin{equation}
= \frac{1}{2\opl}\frac{g_{\rm eff}^2}{3\pi}\int d\omega_{\gamma'} \frac{ \sqrt{\og^2-\mgamma^2}}{\opl}\delta\left(\omega_{\gamma'}-\frac{\omega^2+ \mgamma^2 - \mphi^2}{2 \omega} \right)
\end{equation}
\begin{equation}
=\frac{1}{2\opl^2}\frac{g_{\rm eff}^2}{3\pi}\sqrt{\left(\opl^2-(\mphi + \mgamma)^2\right)\left(\opl^2-(\mphi - \mgamma)^2\right)}.
\end{equation}
Therefore, the decay rate in the rest frame becomes 
\beq \label{decayrestframe}
\Gamma_{\gamma^*\rightarrow \phi\gamma'}=\frac{1}{3}\frac{\Gg^2}{32\pi}\frac{1}{\opl^3}\left[\left(\opl^2-(\mphi + \mgamma)^2\right)\left(\opl^2-(\mphi - \mgamma)^2\right)\right]^{\frac{3}{2}}.\eeq
Finally, applying the boost factor, we find the decay rate in the plasma frame to be
\beq \label{finaldecayrate}
\Gamma_{\gamma^*\rightarrow \phi\gamma'}=\frac{1}{3}\frac{\Gg^2}{32\pi}\frac{1}{\o\,\opl^2}\left[\left(\opl^2-(\mphi + \mgamma)^2\right)\left(\opl^2-(\mphi - \mgamma)^2\right)\right]^{\frac{3}{2}}.
\eeq
In the massless WISPs limit, our result is
\beq 
\Gamma_{\gamma^*\rightarrow \phi\gamma'}=\frac{1}{3}\frac{\Gg^2}{32\pi}\frac{\opl^4}{\o},
\eeq
in agreement with the one found in \cite{Kalashev:2018bra}.

\section{Perturbative solutions to the second-order equations}{\label{app:LSW}}
In this section we show a detailed calculation of the  second-order differential equations (\ref{eom_1}) and (\ref{eom_2}), and the results for the probability of observing a photon in the regeneration cavity of an LSW experiment. 

Recalling that at the end we are working with the component of the photon field that couples to the dark matter background, according to Eqs.~(\ref{eq:new_eom1})-(\ref{eq:new_eom2}), we assume solutions of the form
\begin{align*}
\mathcal A&=\mathcal A^{(0)} + \mathcal A^{(1)} + \mathcal A^{(2)}
+... \nonumber \\
\phi&=\phi ^{(0)} + \phi ^{(1)} + \phi ^{(2)} +...\, .
\end{align*}
Here the superscript labels the order of the solution, and we assume the amplitude of the fields gets smaller as we increase the order. 

At zeroth order, there is only an incident photon plane wave. We will assume that it propagates along the $x$-axis, with frequency $\omega=|{\bf{ k}}|=k$, and thus 
\begin{equation}
\mathcal A^{(0)}=\mathcal A_0 e^{i k(x-t)}.
\end{equation}
This wave is a source for ALPs, in Eq.~\eqref{app:eq:eom2}.
For the first order ALP equation we then get
\begin{equation}
(\Box + m_{\phi}^2 )\phi^{(1)}=4\eta \sin\theta\, \partial_x \mathcal A_z^{(0)} \cos(\mgamma t).
\end{equation} 
With the general solution given by
\begin{equation}
\phi^{(1)}=4i\eta \sin\theta\,  A_0 k  \int dx' dt' G(x-x',t-t')  \cos(m_{\g'}t') 
e^{ik(x'-t')},\label{eq:axion_1ordera}
\end{equation}
where the {Green's} function is 
\begin{equation}
G(x,t)=-\Theta(t)\left( \frac{i}{4\pi}\right) \int \frac{dq
}{\omega_\f(q)}\left(e^{iqx}e^{i\omega_\f (q)t} - e^{iqx}e^{-i\omega_\f(q)t}\right).
\end{equation}
Here $\Theta$ is the Heaviside {function. Moreover, we recall} $\omega_\f(q)=\sqrt{q^2+m^2_{\phi}}$.
First we focus on the time integral 
\begin{equation}
T=\frac{1}{2}\int_{-\infty}^{\infty} dt' \Theta(t-t')\left( e^{im_{\gamma'}t'}+e^{-im_{\gamma'}t'}\right)\left( e^{i \omega_\f (t-t')}-e^{-i\omega_\f (t-t')}\right)e^{-ik t'}.
\end{equation}
{The solution is composed of two resonant parts,}
\begin{equation}
T(q, t)=\frac{e^{i\Omega_- t}}{2i}\left(\frac{2\omega_\f(q) }{ \Omega_-^2 - \omega_{\f}^2(q)}\right) + 
\frac{e^{-i\Omega_+ t}}{2i}\left(\frac{2\omega_\f(q) }{ \Omega_+^2 - \omega_{\f}^2(q)}\right)\equiv T_-(q,t) + T_+(q,t),
\end{equation}
where we have defined $\Omega_\pm = m_{\gamma'} \pm k$.
Next we integrate over the momentum. Let us consider the expression
\begin{eqnarray}
Q_\pm&=&\int_{-\infty}^{\infty} \frac{dq}{2\pi \omega_\f (q)}T_\pm(t,q) e^{iq(x-x')} 
\\\nonumber
&=& e^{\mp i\Omega_\pm t}\int^{\infty}_{-\infty} \frac{dq}{2\pi}\frac{e^{iq(x-x')}}{\Omega_\pm^2 - \omega_{\f}^2(q)} = e^{\mp i\Omega_\pm t}\int^{\infty}_{-\infty} \frac{dq}{2\pi}\frac{e^{iq(x-x')}}{\left( q+\sqrt{\Omega_\pm^2-m_{\phi}^2}\right)\left(q-\sqrt{\Omega_\pm^2-m_{\phi}^2}\right)}.
\end{eqnarray}
This integral can be performed {using boundary conditions such that $\mathcal{A}^{(0)}$ vanishes in the infinite past, and therefore} adding to $k$ a small positive imaginary part $k \rightarrow k + i\varepsilon$.  The poles are located at
\be
q_\pm = \sqrt{\Omega_\pm^2-m_{\phi}^2} \cong  \sqrt{\Omega_\pm^2-m_{\phi}^2} + i \varepsilon \frac{ (k \pm \mgamma) }{\sqrt{\Omega_\pm^2-m_{\phi}^2}}\equiv  \sqrt{\Omega_\pm^2-m_{\phi}^2} + i \delta_{\pm}.
\ee
The contour closes either in the upper or lower half plane for the cases $x-x'>0$, and $x-x'<0$,  respectively. We are interested in the former case,  resulting in
\beq 
Q_\pm=i\frac{e^{\mp i\Omega_{\pm} t}}{2q_\pm} e^{iq_{\pm} (x-x')}.
\eeq
Using Eq.~\eqref{eq:axion_1ordera}, we find
\bea \label{phi1}
\phi^{(1)}&=&\frac{2\eta\sin\theta \mathcal A_0 k}{i} \int_0^L dx' e^{ikx'}(Q_- + Q_+)\\
&=&{\eta\sin\theta\, k L} \left( \frac{e^{i(\Omega_- t + q_- x)}}{q_-}  \Phi(\kappa_-L) e^{i\kappa_-L/2} + 
\frac{e^{i(-\Omega_+ t + q_+ x)}}{q_+} \Phi(\kappa_+L) e^{i\kappa_+L/2} \right),
\eea
where 
\be
\Phi(X) \equiv \frac{\sin \left(X/2\right)}{X/2}
\label{app:eq:form_factor1}
\ee
and
\be
\kappa_\pm=k-q_\pm. \label{kappa}
\ee
Now, we use this first order solution for the ALP field as a source for photons in the second cavity of the LSW set up. Inserting it into the right hand side of Eq.~(\ref{app:eq:eom2}), we get
\beq
\mathcal A^{(2)}(x,t)=-2\eta\sin\theta\int dx' dt' \ \Theta(t-t'-(x-x')) \ \partial_{x'} \phi^{(1)}(x',t') \, \cos{(\mgamma t')}.
\eeq
From the above expression, we have two contributions, which can be written as
\bea
\mathcal A_{\pm}^{(2)} &=& -i\eta^2 \sin^2\theta \mathcal A_0\, k L\int dx' dt' \ \Theta(t-t'-(x-x'))\nonumber \\
&&\quad\times\left \{   \Phi(\kappa_\pm L) e^{iq_\pm x'}e^{i(\mp \Omega_\pm+\mgamma)t'} + \Phi(\kappa_\pm L) e^{iq_\pm x'}e^{i(\mp \Omega_\pm -  \mgamma)t'}
\right \} e^{i\kappa_\pm\frac{L}{2}}.
\eea
Performing first the time integral,
\bea
&&\!\!\!\!\!\!\!\!\!\!\!\!\!\!\!\!\!\!\!\!\!\!\!\!\int dt'  \Theta(t-t'-(x-x')) \left\{ e^{\mp i(2\mgamma\pm k)t'} +  e^{- ikt'}\right\}\Phi(\kappa_\pm L)e^{iq_{\pm}x'}e^{i(k-q_\pm)\frac{L}{2}} \nonumber \\
&&\quad=\left\{\pm i \frac{e^{\mp i(2\mgamma \pm k)t}e^{\pm i(2\mgamma\pm k)(x-x')}}{(2\mgamma\pm k)} + i\frac{e^{ -ikt}e^{i k(x-x')}}{k}
\right\}\Phi(\kappa_\pm L)e^{iq_{\pm}x'}e^{i\kappa_\pm\frac{L}{2}},
\eea
{and then the} integral in $x'$ we find 
\bea \label{a2}
\mathcal A_{\pm}^{(2)}
={\eta^2 \sin^2\theta \mathcal{A}_0 k L{{^2}}} \left\{ \pm \frac{e^{i(k \pm 2\mgamma)(x-t)}}{( k \pm 2\mgamma)} \Phi\left(\chi_\pm L\right) e^{\mp i \mgamma L} - \frac{e^{ ik(x-t)}}{k} \Phi(\kappa_\pm L) \right \} \Phi(\kappa_\pm L)
\label{app:eq:A_second_order}
\eea
where $\chi_-$ and $\chi_+$ are defined as 
\be
\chi_\pm=-k\mp2\mgamma+q_\pm \label{chi}
\ee

Therefore, the second order solution for the photon field in the regeneration cavity is $\mathcal A^{(2)}(x,t)=\mathcal A_+(x,t)+\mathcal A_-(x,t)$.

\subsection{Probability in an LSW experiment}
From the second order amplitude result, Eq.~\eqref{app:eq:A_second_order}, we obtain the probability to observe a photon in an LSW-type experiment, (now writing in terms of the incoming photon frequency $\omega=k$)
\begin{eqnarray}
\label{eq:prob}
P_{\gamma \gamma} &=& \frac{\Gg^4 \rho_{\rm CDM}^2L^4\sin^4\theta}{64}\bigg(
\frac{\o^2}{(\o-2\mgamma)^2}\,\Phi(\kappa_-L)^2\Phi(\chi_-L)^2+\frac{\o^2}{(\o+2\mgamma)^2}\, \Phi(\kappa_+L)^2\Phi(\chi_+L)^2 \nonumber
\\
& & + \Phi(\kappa_-L)^4 +\Phi(\kappa_+L)^4 +2\Phi(\kappa_-L)^2 \Phi(\kappa_+L)^2 \bigg), 
\end{eqnarray}
where $\rho_{\rm CDM }=0.3$~GeV~cm$^{-3}$ is the local dark matter density.
Let us note that the calculation is only valid when $\mgamma\neq\o/2$. In this limit, a zero frequency photon emerges in the regeneration cavity, and the term $\frac{\o^2}{(\o- 2\mgamma)^2}$ would seem to have a resonance there, but {this is an unphysical} result.

In the limit where the masses are much smaller than the frequency of the incoming $(\mgamma,\mphi) \ll k$, we have
\be
P_{\gamma \gamma}=\frac{\Gg^4 \rho_{\rm CDM}^2 {L^4}\sin^4\theta }{32}\left(\Phi(\kappa_-L)^4 +\Phi(\kappa_+L)^4 +\Phi(\kappa_-L)^2 \Phi(\kappa_+L)^2  \right).
\ee

Finally, let us note that the probability, Eq.~\eqref{eq:prob}, includes the probability to have photons with $\omega$ but also $\omega\pm 2m_{\gamma'}$. In an experiment where a high-finesse cavity is also employed on the regeneration side~\cite{Hoogeveen:1990vq,Mueller:2009wt,Bahre:2013ywa,Graham:2015ouw}, photons with different frequency are likely not to be enhanced by the cavity if they are outside the bandwidth of the cavity. The corresponding contributions (first line in Eq.~\eqref{eq:prob}) would therefore have to be removed from the probability.
In Figs.~\ref{fig:overview_massless} and \ref{fig:overview_10mhp} we have employed this as a conservative approximation over the entire mass range of the ALPS II line.

\subsection{Birefringence and dichroism}

From the perturbative approach developed at the beginning of this appendix, we also can find effects that are relevant in optical polarisation experiments measuring birefringence and dichroism. Computing the second order correction for the electromagnetic field that takes place in the first cavity, we get
\be
\mathcal{A}^{(2)}(x,t)=\eta^2\sin(\theta)^2x^2\mathcal{A}_0e^{ik(x-t)}\sum_{+,-}\left(F(0,\kappa_\pm,x)+e^{2i\mgamma t}F(\mgamma,\kappa_\pm,x)\right) \label{A2optical}
\ee
where
\bea
F(\mgamma,\kappa_\pm,x) &=& \frac{\kappa_\pm}{ \kappa_\pm-2\mgamma}\left(\frac{\mgamma}{\kappa_\pm}\Phi(2\mgamma x)^2-\frac{1}{2}\Phi(\kappa_\pm x)^2
\right. \nonumber
\\
&&\left.+\frac{i}{\kappa_\pm x}\left(\Phi(4\mgamma x)-\Phi(2\kappa_\pm x)\right)\right). \label{F}
\eea

In high finesse cavities the term oscillating with the extra factor $e^{2i\mgamma t}$ is often not in the resonance width. Considering this case and dropping the corresponding term\footnote{Averaging the measurements over a sufficiently long time should yield a similar result.} we can extract from~\eqref{A2optical} the ellipticity produced by the phase change on $\mathcal{A}$
\be
\varepsilon(x)=\eta^2\sin(\theta)^2x^2{\sin(2\Theta)}\left(\frac{1}{2\kappa_+ x}\left(1-\Phi(2\kappa_+x)\right)+\frac{1}{2\kappa_-x}\left(1-\Phi(2\kappa_-x)\right)\right), \label{eq:elip}
\ee
where $\Theta$ is the initial angle formed by the laser polarisation and the DM electric field. We can also find the change in the polarisation angle. It is given by
\be
\delta\Theta(x)=\frac{1}{2}\eta^2\sin(\theta)^2x^2\frac{\sin(2\Theta)}{2}\left(\Phi(\kappa_+x)^2+\Phi(\kappa_-x)^2\right). \label{eq:polangle}
\ee

For the zero hidden photon mass case, which contradicts our HP dark matter assumption, we can not neglect that term and we get the usual result for the axion-photon oscillation in an static magnetic field $B_0$ having done the replacement $\Gg E_0^\prime \rightarrow g_{\phi\gamma\gamma}B_0$.

\bibliographystyle{utphys}
\bibliography{bibliography.bib}

\end{document}